\documentclass[aps,pre,twocolumn,superscriptaddress,showpacs]{revtex4-1}

\usepackage{epsfig}
\usepackage{amssymb}
\usepackage{color}
\usepackage{amsbsy}
\usepackage{amsmath}

\begin{document}

\title{Rotational and translational diffusion in an interacting active dumbbell system}


\author{Leticia F. Cugliandolo}
\affiliation{Sorbonne Universit\'es, Universit\'e Pierre et Marie Curie - Paris VI, \\
Laboratoire de
Physique Th\'eorique et Hautes \'Energies, \\
4 Place Jussieu, 75252 Paris Cedex 05,
France}

\author{Giuseppe Gonnella}
\email{leticia@lpthe.jussieu.fr}
\affiliation{Dipartimento di Fisica, Universit\`a di Bari {\rm and}  \\
INFN, Sezione di Bari, via Amendola 173, Bari, I-70126,
Italy}
\email{gonnella@ba.infn.it}

\author{Antonio Suma}
\affiliation{SISSA - Scuola Internazionale Superiore di Studi Avanzati,\\
Via Bonomea 265, 34136 Trieste 
Italy}
\email{antonio.suma@gmail.com}

\begin{abstract}
We study the dynamical properties of a two-dimensional ensemble 
of self--propelled dumbbells with  only repulsive interactions. This model undergoes a 
phase transition between a homogeneous and a segregated phase and we focus on the 
former. We analyse the translational and rotational  mean square displacements  in terms of the 
P\'eclet number, describing the relative role of active forces and thermal fluctuations,  
and of particle density. We find that the four distinct regimes of the translational mean square displacement of the 
single active dumbbell survive at finite density for parameters that lead to a separation of time-scales.
We establish the P\'eclet number and density dependence of the diffusion constant in the 
last diffusive regime. We prove that the ratio between the  diffusion constant and its value for the single dumbbell
depends on temperature and active force only through the P\'eclet number at all densities explored.
We also study the  rotational mean square displacement  proving the existence of a rich 
behavior with intermediate  regimes only appearing at finite density. 
The ratio of the rotational late-time diffusion constant and  its vanishing density limit
depends on the P\'eclet number and density only. At low P\'eclet number it is a monotonically decreasing function of density. At
high P\'eclet number it first increases to reach a maximum and next decreases as a function of density.
We interpret the latter result advocating the presence of large-scale fluctuations  close 
to the transition, at  large enough density,  that favour coherent rotation inhibiting, however,  rotational motion for even larger packing fractions.
\end{abstract}


\maketitle

\setlength{\textfloatsep}{10pt} 
\setlength{\intextsep}{10pt}


\section{Introduction}
\label{sec:introduction}

{Active matter includes different kinds of self-driven systems  which live, or function, far from thermodynamic equilibrium, by   
continuously converting internal energy sources  into work or movement [1,2].}
Nature offers many examples {of this kind of 
 condensed matter}, at very different scales: the cytoskeleton, 
bacterial colonies and  algae suspensions, bird flocks and schools of
fish {are just some} among others~\cite{Toner05,Fletcher09,menon,Ramaswamy10,Cates12,Romanczuk12,Vicsek12,Marchetti13,Marenduzzo14,gonnella2015rev}. 
Self-propelled units  can also be artificially realized in the laboratory in 
many different ways, for example, by surface treatment of colloidal particles~\cite{Walther}. 

Active matter is inherently  out of equilibrium and exhibits non-trivial properties
that have no analogue in passive, equilibrium  materials. For example, large scale coherent motion and 
self-organised dynamic structures have been observed in colonies of bacteria in the absence of any attractive 
interaction~\cite{Mendelson99,wu,Dombrowski04,Hernandez05,Riedel05,Sokolov07,Zhang09}.
{In addition,} a phase separation into an aggregate and a gas-like phase has been found in 
theoretical models~\cite{Tailleur08,Fily12,Fily14,Redner13,Stenhammer13,Suma13,Suma14,Levis-Berthier,Stenhammer14}
and, recently,  also in experiments~\cite{Buttinoni13b}  {on}  suspensions of self-motile particles only subject to steric interactions.  
   
The dynamical properties of a suspension are significantly affected by self-propulsion.
For instance, the comparison of the diffusion constant $D$ of a dilute solution of passive spherical colloids with 
the one of run-and-tumble bacteria shows that the Stokes-Einstein
formula and the fluctuation-dissipation theorem do not hold for the active 
system~\cite{wu,Palacci10}. Indeed,  
assuming the Stokes-Einstein relation, $D$ is given by 
$D=k_B T / (3 \pi \eta \sigma)$ where $T$ is the temperature, $\eta$ the fluid viscosity, and 
$\sigma$ the diameter of the colloids.  
The diffusion coefficient in a three-dimensional system of run-and-tumble bacteria~\cite{Cates12} is
evaluated as $D_A \approx l^2_{run}/(6\tau)$ where $\tau$ is the duration of each run and 
$l_{run}$ its length. Using $\tau \approx 1$ s,  
$l_{run} \approx 10-30 \ \mu$m, water viscosity $\eta \approx 10^{-3}$ Pa s, and $\sigma \approx 1 \ \mu$m, 
one finds $D_A/D \approx 10^2$. 

This {simple order-of-magnitude} argument suggests  to analyse  more carefully how the diffusive 
behaviour is affected by self-propulsion.  Actually, several experimental studies addressed this question.
Wu and Libchaber considered the mean-square displacement of {\it passive tracers} coupled to a dilute suspension of 
Escherichia coli and found that an initial  super-diffusive behaviour  crosses over to normal diffusion at late times~\cite{wu}. 
The super-diffusive behaviour was interpreted as being due to the presence of coherent structures in the bacterial bath.
A similar study was carried on by Leptos {\it et al.} on a suspension of algae. These authors found a linear time-dependence 
of the passive tracer's mean-square displacement at all measured {time-lags}~\cite{Leptos09}. 
The tracer's diffusive constant was found to depend 
linearly on the density of swimmers~\cite{Leptos09} while,  reducing the   dimensionality {to a film-like geometry}, 
the density-dependence 
was enhanced to the power 3/2~\cite{Kurtuldu11}. Results on the  diffusion coefficient of tracers in contact 
with bacterial suspensions with hydrodynamics playing  a relevant role were given 
in~\cite{Kasyap14,Pushkin14,Morozov14}.

Other studies focused on the mean-square displacement of the {\it active particles} themselves.
A linear dependence of the diffusion constant of the active swimmers 
on the so-called active flux (active swimmers density times their mean velocity)  
was found for different kinds of swimmers in contact with a solid wall~\cite{Mino11}.
The simulations of Hern\'andez-Ort\'{\i}z {\it et al.}  show ballistic behaviour crossing over to  normal diffusion 
for the swimmer and passive tracer particles though with different density-dependence of the diffusion constant
in the latter regime. The {swimmer's } diffusion constant decreases with density at low swimmer density while it increases at large values;
the diffusion constant of the passive tracer has, instead, a monotonic dependence with density and it 
consistently increases with it~\cite{Hernandez05}.  The Lattice Boltzmann study in~\cite{Llopis06} also shows
a cross-over between ballistic and diffusive behaviour at relative high self-propelled particle density and 
a super-diffusive regime associated to the formation of large scale clusters at low densities.

A detailed study of diffusion properties in models for self-propelled particles at different densities 
 for the interesting cases where self-propulsion has been recognised 
to be an interaction capable of producing a phase transition is lacking in the literature.
{Some} results for the case of self-propelled polar disks have been presented in~\cite{Fily12}.
The paper by Gr\'egoire {\it et al.}~\cite{Gregoire01}, 
who considered a model with particle interaction {that favors alignment} {\it \`a la Vicsek}, 
{can also be mentioned in this respect. These authors}
interpreted  the super-diffusive behaviour 
of~\cite{wu} in terms of  the  cross-over found in their model close to the critical point. 

Swimmers typically have elongated shape. We have therefore  decided to study
the diffusive properties of rotational and translational degrees of freedom in a system of active dumbbells.
This model was introduced  in~\cite{valeriani2011colloids} to describe the experimental behaviour 
of a bacterial bath coupled to colloidal tracers. A first study of its phase diagram appeared  in~\cite{Suma13,Suma14}
where it was shown that for certain (relatively high) densities and active forces the system phase separates
into dense and loose spatial regions.
A later {work}~\cite{Suma14b}  focused on
the dumbbell effective temperature defined 
in terms of a fluctuation-dissipation relation~\cite{cugl:review}. 
However, a detailed analysis of the translational and rotational mean-square displacements
in the full range of time-delays and varying the three more relevant parameters 
(temperature, activity and density) simultaneously, 
was not performed yet. We present {such a complete analysis in the homogeneous phase 
here}.

The paper is organised as follows. In Section~\ref{sec:model-dumbbells}  the dumbbell model is reviewed. 
The diffusion behavior of a single dumbbell  is described  in Section~\ref{sec:single-dumbbell}.
In Section~\ref{sec:numerical}  the numerical results 
for the interacting active system are presented. The phase diagram 
 is analysed in terms of the P\'eclet number. This is a preliminary step needed to fix the region
of the parameter space to be  considered for the  measurements of the diffusive properties. 
Then the translational and rotational mean-square displacements {in the homogeneous phase} 
are studied in all their dynamic regimes. Special emphasis is 
put on the analysis of the parameter dependence of the diffusion coefficient in the asymptotic limit.
A discussion will complete the paper in Section~\ref{sec:conclusions}.

\section{The model}
\label{sec:model-dumbbells}

A dumbbell is a diatomic molecule formed by two spherical colloids with diameter $\sigma_{\rm d}$ 
and mass $m_{\rm d}$ linked together. The atomic  positions are noted ${\mathbf r}_1$ and ${\mathbf r}_2$
in a Cartesian system of coordinates fixed to the laboratory.  
The colloids are subject to internal and external forces.

Typically, one assumes that there is an elastic link between the colloids modeled by 
the finite extensible non-linear elastic force
\begin{equation}
{\mathbf F}_{\rm fene} = - \frac{k {\mathbf r}}{1- (r^2/r_0^2)} 
\end{equation}
with $k>0$.
The denominator ensures that the spheres cannot go beyond the distance $r_0$
with $r$ the distance between their centres of mass.
An additional repulsive force is added to ensure that the two colloids do not overlap. This 
is the Weeks-Chandler-Anderson (WCA) potential~\cite{Weeks} 
\begin{eqnarray}
\label{eq:WCA-potential}
V_{\rm wca}( r ) 
&=&
\left\{
\begin{array}{ll}
V_{\rm LJ}( r ) - V_{\rm LJ}(r_c) & \qquad r<r_c
\\
0 & \qquad r > r_c
\end{array}
\right.
\end{eqnarray}
with 
\begin{equation}
V_{\rm LJ}(  r  ) = 4\epsilon \left[ \left( \frac{\sigma_{\rm d}}{r} \right)^{12} - \left( \frac{\sigma_{\rm d}}{r} \right)^{6}\right]
\; ,
\end{equation}
where $\epsilon$ is an energy scale and $r_c$ is the minimum of the Lennard-Jones potential, $r_c=2^{1/6} \sigma_{\rm d}$.

The active forces are polar. They  act along the main molecular axis  $\hat {\mathbf n}$, are constant in modulus
{pointing in the same direction}
for the two 
spheres belonging to the same molecule {     
~\footnote{  In a system with momentum conservation the total force on a neutrally buoyant swimmer should indeed be zero. However Brownian dynamics theories and  simulations neglect fluid-mediated interactions so the only way to propel a particle is to apply a force along its direction.}}, 
and read
\begin{equation}
{\mathbf F}_{\rm act} = F_{\rm act} \ \hat {\mathbf n}
\; . 
\end{equation}

We take the interaction between the spheres in different dumbbells to be purely repulsive
{and of the same WCA form as for the two colloids composing  one dumbbell.}
 
The dynamic equations for one dumbbell  are
\begin{eqnarray}
m_d\ddot{{\mathbf r}}_{i}(t)  &=&  -\gamma \dot{{\mathbf r}}_{i}(t)+
 {\mathbf F}_{\rm fene}({\mathbf r}_{i, i+1})+
 {\boldsymbol \eta}_{i} 
  \nonumber\\
 &&
 - \!
 \sum_{\substack{
            j=0\\
            j \neq i}}^{2N} 
 \frac{\partial V_{\rm wca}^{ij}}{\partial  r_{ij}}
 \frac{{\mathbf r}_{ij}}{ r_{ij}}+ {{\mathbf F}_{\rm act}}_i
\; , 
\label{eqdumbattcoll}
\\
 m_d\ddot{{\mathbf r}}_{i+1}(t) &=& 
 -\gamma \dot{{\mathbf r}}_{i+1}(t)-
 {\mathbf F}_{\rm fene}({\mathbf r}_{i,i+1})+
 {\boldsymbol \eta}_{i+1}
 \nonumber\\
&& - \! \sum_{\substack{
            j=0\\
            j \neq i+1}}^{2N} 
\!\!  \frac{\partial V_{\rm wca}^{i+1,j}}{\partial r_{i+1,j}}
 \frac{{\mathbf r}_{i+1,j}}{r_{i+1,j}}+{{\mathbf F}_{\rm act}}_i  
 \; , 
 \;\;\;\;
\end{eqnarray}
with $i=1,3,...2N-1$,  ${\mathbf r}_{ij} = {\mathbf r}_i - {\mathbf r}_j$, $r_{ij} = |{\mathbf r}_{ij}|$
 and $V_{\rm wca}^{ij} \equiv
V_{\rm wca}(r_{ij})$ with $V_{\rm wca}$ defined in Eq.~(\ref{eq:WCA-potential}).
Once the active force is attached to a molecule 
a sense of back and forth atoms is attributed to them; ${\mathbf F}_{\rm act}$ is  directed 
from the $i$th colloid (tail) to the $i+1$th colloid (head). The active forces are  applied to all molecules
in the sample during all their dynamic evolution. 
${\mathbf F}_{\rm act}$ changes direction 
together with the molecule's rotation. 

The coupling to the thermal bath is modelled as usual, with a friction and a noise 
term added to the equation of motion.
$\gamma$ is the friction coefficient and {we do not distinguish friction along the main molecular axis and transverse to it, 
as done in some publications~\cite{Baskaran10}.}
The noise ${\boldsymbol \eta}$ is a  Gaussian random variable with 
\begin{eqnarray}
\langle \eta_{ia}(t) \rangle &=& 0 \; , 
\label{eq:noise-ave}\\
\langle \eta_{ia}(t) \eta_{jb}(t') \rangle &=& 2 \gamma k_BT \delta_{ij} \delta_{ab} \delta(t-t')
\; ,
\label{eq:noise-corr}
\end{eqnarray}
with $k_B$ the Boltzmann constant and $T$ the temperature of the equilibrium environment in which the 
dumbbells move. $a$ and $b$ label the coordinates in $d$ dimensional space. An effective 
rotational motion is generated by the random torque due to the white noise acting independently on the two beads.
{
We consider initial conditions at time $t=0$ such that the initial angle $\theta_0 = \theta(0)$, randomly distributed
between $[-\pi,\pi]$, 
has zero mean $[ \theta_0 ]_{ic} = 0$.}

The surface fraction is 
\begin{equation}
\phi=N \ \frac{S_{\rm d}}{S} 
\;  
\end{equation}
with $S_{\rm d}$ the area occupied by an individual dumbbell, $S$ the total area of the box where the dumbbells move and $N$ their total number. 
The spring is supposed to be massless and void of surface. Therefore, in  $d=2$, $S_{\rm d} = \pi \sigma_{\rm d}^2/2$.  We impose 
periodic boundary conditions on the two directions. 


The P\'eclet number, ${\rm Pe}$, is a dimensionless ratio between the  
advective transport rate and the diffusive transport rate.
For particle flow one defines it as 
$
{\rm Pe} = Lv/{D} 
$,
with $L$ a typical length, $v$ a typical velocity, and $D$ a typical diffusion constant. We choose $L \to \sigma_{\rm d}$, $v\to F_{\rm act}/\gamma$ and 
$D\to D^{\rm pd}_{\rm cm}= k_BT/(2\gamma)$ of the passive dumbbell to be derived below; then,
\begin{equation}
{\rm Pe} = \frac{2\sigma_{\rm d} F_{\rm act}}{k_BT}
\; . 
\label{eq:Peclet}
\end{equation}
{This parameter is also a measure of the ratio between the work done by the active force in 
translating  the center of mass of the molecule  by a distance  of $2 \sigma_{\rm d}$,
and the thermal energy scale.}
Another important parameter is the active Reynolds number 
\begin{equation}
{\rm Re}_{\rm act} 
= \frac{m_{\rm d} F_{\rm act} }{\sigma_{\rm d} \gamma^2} 
\; , 
\label{eq:Reynolds}
\end{equation}
defined in analogy with the usual hydrodynamic Reynolds number ${\rm Re}=  L v /\nu$,
where $\nu $ is the kinematic viscosity of a given fluid, representing the ratio between inertial
and viscous forces. Here we set
 $L \to \sigma_{\rm d}$, $v\to F_{\rm act}/\gamma$ and 
$\nu \to \gamma \sigma_{\rm d}^2 /m_{\rm d}$.

\section{A single dumbbell}
\label{sec:single-dumbbell}

{Before studying the interacting problem with  numerical simulations in Sec.~\ref{sec:numerical}, 
we derive analytically the translational and rotational mean-square displacements of the single 
dumbbell.}

The equation of motion for the position of the centre of mass, ${\mathbf r}_{\rm cm}= ( {\mathbf r}_1 + {\mathbf r}_2)/2$,
of a single dumbbell is
\begin{equation}
2 m_{\rm d} \ddot {\mathbf r}_{\rm cm}(t) = - 2 \gamma \dot {\mathbf r}_{\rm cm}(t) + 2 {\mathbf F}_{\rm act}(t) + {\boldsymbol \xi}(t)
\label{eq:Langevin-cm}
\end{equation}
with the new noise ${\boldsymbol \xi}(t) \equiv {\boldsymbol \eta}_1(t) + {\boldsymbol \eta}_2(t)$  with vanishing
average, $\langle \xi_a(t) \rangle =0$, and correlation 
\begin{equation}
\langle \xi_a(t) \xi_b(t') \rangle = 4\gamma k_BT \ \delta_{ab} \delta(t-t')
\label{eq:corr_rum}
\; . 
\end{equation}
This is the Langevin equation of a point-like particle with mass $2m_{\rm d}$, under a force $2{\mathbf F}_{\rm act}$, and  in contact with a bath with friction coefficient 
$2\gamma$ at temperature $T$. 

The equation of motion for the relative position of the two monomers, ${\mathbf r} = {\mathbf r}_1-{\mathbf r}_2$,  is 
\begin{equation}
m_{\rm d} \ddot {\mathbf r}(t) = {-} \gamma \dot {\mathbf r}(t) + 2 {\mathbf F}_{\rm int}(t)  + {\boldsymbol \zeta}(t)
\label{eq:Langevin-r12}
\end{equation}
with the new noise ${\boldsymbol \zeta}(t) =  {\boldsymbol \eta}_1(t) - {\boldsymbol \eta}_2(t)$ having zero average,
$\langle \zeta_a(t) \rangle =0$, and correlation
\begin{equation}
\langle \zeta_a(t) \zeta_b(t') \rangle = 4\gamma k_BT \ \delta_{ab} \delta(t-t')
\; . 
\end{equation}

Note that the noises ${\boldsymbol \xi}$ and ${\boldsymbol \zeta}$ are independent, 
$\langle \xi_a(t) \zeta_b(t') \rangle = 0$, for all $a, b$ at all times. ${\mathbf F}_{\rm int} $ includes  the elastic and 
repulsive  forces internal to the single dumbbell.

Equation~(\ref{eq:Langevin-r12})  controls the molecule's elongation and its rotational motion
while Eq.~(\ref{eq:Langevin-cm}) determines the translational properties of the dumbbell.
{The internal force ${\mathbf F}_{\rm int}$ affects the elongation of the molecule
while the thermal noise adds fluctuations to it but, more importantly, it applies an effective torque 
and induces rotations.}
Equations~(\ref{eq:Langevin-cm}) and (\ref{eq:Langevin-r12}) are coupled by the fact that ${\mathbf F}_{\rm act}$ 
acts along the axis of the molecule, the orientation of which changes in time {in the presence of thermal 
fluctuations}.

\subsection{Elongation and rotation}

Let us call $\hat {\mathbf u}_\parallel$ the instantaneous unit vector pointing from monomer 1 to monomer 2 along 
the axis of the molecule, $\theta$ the angle between $\hat {\mathbf u}_\parallel$ and an axis fixed to the laboratory, and 
$\hat {\mathbf u}_\perp$ a unit vector that is perpendicular to $\hat {\mathbf u}_\parallel$ at all times. 
Using $\dot {\mathbf r} = \dot r \hat {\mathbf u}_\parallel + r \dot {\hat {\mathbf u}}_\parallel$, with $r$ the modulus of ${\mathbf r}$, 
$\dot{\hat {\mathbf u}}_{\parallel} = \dot \theta \hat {\mathbf u}_\perp $  
{ and } 
$\dot{\hat {\mathbf u}}_{\perp}     = -\dot \theta \hat {\mathbf u}_{\parallel}$ (note that we use here 
the Stratonovich discretisation scheme of stochastic differential equations~\cite{Oksendhal} and we are thus 
entitled to apply the usual rules of calculus)~{ ~\footnote{The use 
of Stratonovich calculation is quite natural in this context, as stressed by van Kampen and others \cite{van1981ito}, as  for most of physical problems. We can also observe that, following the Ito approach~\cite{Gardiner},  the equations of motion for the polar coordinates
 can be written, neglecting the inertial contribution, as 
$\gamma \dot r = 2 F_{\rm int} + \frac{2k_BT}{r}+\zeta_r$, 
$\dot \theta = \frac{1}{\gamma r} \zeta_\theta$ with $\zeta_r$,  $\zeta_\theta$ Gaussian white noises satisfying  the same properties as $\zeta_x$, $\zeta_y$ of Eq.~(\ref{eq:corr_rum}). This  set of equations give the same dynamical equations for the momenta Eqs. (\ref{eq:rrr}-\ref{eq:thetptoquadr}) resulting from Stratonovich approach. }}, one has 
\begin{eqnarray}
m_{\rm d} (\ddot r - r {\dot \theta}^2) &=& -\gamma \dot r + 2 F_{\rm int} + \zeta_\parallel
\; , 
\label{eq:Langevin-rtheta1}
\\
m_{\rm d} (2 \dot r \dot \theta + r \ddot \theta) &=&  - \gamma r \dot \theta + \zeta_\perp
\; , 
\label{eq:Lanvegin-rtheta2}
\end{eqnarray}
where we decomposed the noise into the parallel and perpendicular directions, 
${\boldsymbol \zeta} = \zeta_\parallel \hat {\mathbf u}_\parallel + \zeta_\perp \hat {\mathbf u}_\perp$.
The relations between the unit vectors in the fixed laboratory and the co-moving frame are
given by 
\begin{eqnarray}
\hat {\mathbf u}_x &=& 
\cos\theta \ \hat {\mathbf u}_\parallel - \sin\theta \ \hat {\mathbf u}_\perp
\; , \qquad\quad
\nonumber\\ 
\hat {\mathbf u}_y &=& 
\sin\theta \ \hat {\mathbf u}_\parallel + \cos\theta \ \hat {\mathbf u}_\perp
\; .
\end{eqnarray}
With this, for any noise we write
\begin{eqnarray}
{\boldsymbol \zeta}  \ &=& \
(\zeta_x \ \cos\theta + \zeta_y \ \sin\theta) \ \hat {\mathbf u}_\parallel
\nonumber\\
&&
+
(-\zeta_x \ \sin\theta + \zeta_y \ \cos\theta) \ \hat {\mathbf u}_\perp
\; , 
\end{eqnarray}
and 
\begin{eqnarray}
\zeta_\parallel &=& \zeta_x \ \cos\theta + \zeta_y \ \sin\theta
\; , 
\nonumber\\ 
\qquad\quad
\zeta_\perp &=& -\zeta_x \ \sin\theta + \zeta_y \ \cos\theta
\; . 
\end{eqnarray}

The system of equations (\ref{eq:Langevin-rtheta1})-(\ref{eq:Lanvegin-rtheta2}) for $r$ and $\theta$ 
cannot be solved exactly. We will assume that the 
internal and viscous forces are such that 
 the inertial contributions (all terms proportional to $m_{\rm d}$)
can be neglected. We then have 
\begin{eqnarray}
\gamma \dot r &=& 2 F_{\rm int} + \zeta_x \ \cos\theta + \zeta_y \ \sin\theta
\; , 
\label{eq:r}
\\
\gamma r \dot\theta &=& -\zeta_x \ \sin\theta + \zeta_y \ \cos\theta
\; . 
\label{eq:theta}
\end{eqnarray} 
Putting together $r$ and $\theta$ into a vector   ${\mathbf y}=(r, \theta)$, this set of equations reads
\begin{equation}
\dot y_\alpha = h_\alpha[{\mathbf y}] + g_{\alpha\beta}[{\mathbf y}] \ \zeta_\beta
\end{equation}
where the index $\beta$ is a Cartesian one, 
$\zeta_1=\zeta_x$ and $\zeta_2 = \zeta_y$, 
and the index $\alpha$ yields $y_1=r$ and $y_2=\theta$. The components of the vector ${\mathbf h}$ and the 
matrix  ${\mathbf g}$ can be easily read from Eqs.~(\ref{eq:r}) and (\ref{eq:theta}). In the last term the noise appears multiplying a 
function of the stochastic variable ${\mathbf y}$. 

One can now average Eqs.~(\ref{eq:r}) and (\ref{eq:theta})  over the Cartesian white noise by  
using the rules of Stratonovich stochastic calculus described in~\cite{Oksendhal,Gardiner,Coffey}:
\begin{equation}
\langle g_{\alpha\beta}  \zeta_\beta \rangle = D \langle g_{\nu\beta} \frac{\partial}{\partial y_\nu} g_{\alpha\beta} \rangle
\end{equation}
where all factors are evaluated at the same time and $D=2 \gamma k_BT$. 
The explicit  calculation  yields
\begin{eqnarray}
\gamma \frac{{\rm d} \langle r \rangle} {{\rm d} t} &=& 2 \langle F_{\rm int} \rangle + D \gamma^{-1} \langle r^{-1} \rangle
\label{eq:rrr}
\; , 
\\
\gamma \frac{{\rm d} \langle \theta\rangle}{{\rm d} t} &=& 0
\; . 
\end{eqnarray}
The first equation is independent of $\theta$ but it involves the average of different functions of $r$. 
The second equation implies 
$\langle \theta\rangle =\theta_0 =\theta(0)$ and, as we will take random initial conditions with average
$[\theta_0]_{ic}=0$, then
$[\langle \theta \rangle]_{ic} =0$. The statistics of $\theta$ can be further analysed 
from the equation for the angular variance
\begin{eqnarray}
\gamma  \frac{{\rm d}}{{\rm d}t} \langle \theta^2\rangle =  2D\gamma^{-1}  \langle r^{-2} \rangle
\; . 
\label{eq:thetptoquadr}
\end{eqnarray}
If we assume that $r$ does not fluctuate  around $\sigma_{\rm d}$,
otherwise stated, that the molecule is approximately 
rigid~\footnote{ {Equation~(\ref{eq:rrr}) with the l.h.s. set to zero and the potential parameters that we use 
in the simulations yields $r\approx 0.96 \sigma_{\rm d}$ quite independently of temperature in the range 
$k_BT \in [10^{-5}, 1]$. In the simulations we find that the fluctuations around this value increase weakly with increasing temperature.}}, 
\begin{equation}
r \approx  \sigma_{\rm d} 
\; , 
\end{equation}
this equation implies angular diffusion 
\begin{equation}
\langle \theta^2\rangle = \theta_0^2 + 2 D_R t 
\end{equation}
with the angular diffusion constant
\begin{equation}
\qquad D_R = \frac{D}{\gamma^2 \sigma_{\rm d}^2} = \frac{2 k_BT}{\gamma \sigma_{\rm d}^2}
\label{eq:DR}
\; . 
\end{equation}
The same technique can be used to compute all moments of the angular variable and thus show that it is 
Gaussian distributed within the rigid molecule approximation.

{
It is interesting to  compare our expression for the rotational diffusion constant, that for the sake of clarity we call 
$D_R^{\rm dumb}$ in 
this paragraph, with that for 
self-propelled hard rods, as described in the  Langevin approach by   Baskaran and Marchetti~\cite{Baskaran10}.
 In the model of Ref.~\cite{Baskaran10}  the rotational diffusion constant is given by  $D_R^{\rm rod}= (k_B T m)/(I \gamma)$ 
where $I$ is the moment of inertia calculated along the main axis of the rod and  $\gamma$ is a friction coefficient.
If  we identify the latter $\gamma$ with our friction coefficient,    
and  we  take  for the dumbbell the moment of inertia of two point-like particles with mass $m=m_{\rm d}$ and 
diameter $\sigma_{\rm d}=2R$, $I= 2 m R^2 $, then
our  expression  for the rotational diffusion coefficient coincides with the one in~\cite{Baskaran10}.
Consider now a rod of length $l$ with the same aspect ratio  
($l=4R$, $\sigma=2 R$)
 and the same total mass $2m_{\rm d}$ of the dumbbell. One finds $I =2 m_{\rm d} R^2 A$ with $A \approx 2.94$~\cite{Baskaran10},
 so that $D_R^{\rm dumb} = A D_R^{\rm rod} > D_R^{\rm rod}$. On the other hand, in the limit of a very long rod ($l \gg R$) one has 
 $I= m l^2/6$ and $D_R^{\rm dumb} = (D_R^{\rm rod} l^2) /(12 R^2)$, so that  $D_R^{\rm dumb}  \gg D_R^{\rm rod}$ as expected.  
Therefore the comparison with  the model of Ref~\cite{Baskaran10} suggests
that the rotational diffusion coefficient of a dumbbell is  always larger than the one of a rod in a suspension.}

\subsection{The center of mass}

We now focus on the statistical properties of the centre of mass position and velocity that depend upon the active force. 
One readily solves Eq.~(\ref{eq:Langevin-cm}) 
\begin{eqnarray}
&&
{\mathbf r}_{\rm cm}(t) 
=
\left( {\mathbf r}_0 + \frac{{\mathbf v}_0 m_{\rm d}}{\gamma} \right) 
- 
\frac{{\mathbf v}_0 m_{\rm d}}{\gamma} \ e^{-\frac{\gamma}{m_{\rm d}} t} 
\nonumber\\\label{eq:r-cm}
&&
\quad
+ \frac{1}{2\gamma} \int_0^t \! dt' \  [1-e^{-\frac{\gamma}{m_{\rm d}} (t-t')}]  [2{\mathbf F}_{\rm act}(t') + {\boldsymbol \xi}(t') ]
\; , 
\\
&& {\mathbf v}_{\rm cm}(t) 
=
{\mathbf v}_0 \ e^{-\frac{\gamma}{m_{\rm d}} t} 
\nonumber\\
&& \quad
+ \frac{1}{2m_{\rm d}} \int_0^t \! dt' \  e^{-\frac{\gamma}{m_{\rm d}} (t-t')} \ [2{\mathbf F}_{\rm act}(t') + {\boldsymbol \xi}(t') ]
\; , 
\label{eq:velocity-cm}
\end{eqnarray}
with ${\mathbf r}_0 = {\mathbf r}_{\rm cm}(0)$ and ${\mathbf v}_0 = {\mathbf v}_{\rm cm}(0)$.

From Eq.~(\ref{eq:velocity-cm}) and thanks to 
$[\langle \cos\theta \rangle ]_{ic} = [\langle \sin\theta\rangle]_{ic} = 0$
 one finds $[\langle {\mathbf v}_{\rm cm}\rangle]_{ic} =0$
and, after some long but straightforward integrations, 
\begin{equation}
2 m_{\rm d} [\langle {v^2_{\rm cm}}_x\rangle]_{ic} =  k_BT + \frac{F_{\rm act}^2}{\gamma (t_I^{-1}+ t_a^{-1})}  
\label{eq:velocity-cm-average}
\end{equation}
with 
\begin{eqnarray}
t_I &=& \frac{m_{\rm d}}{\gamma} \; , 
\nonumber\\
t_a &=& D_R^{-1} = \frac{\gamma \sigma_{\rm d}^2}{2k_BT} = \frac{\sigma_{\rm d}^2}{4 D_{\rm cm}^{\rm pd}}
\; , 
\end{eqnarray}
in the long time limit, beyond $t_I$.
{The time scales 
$ t_I$ and $t_a$ are independent of the active force, and they are  the usual inertial time 
 and a characteristic time associated to  rotational 
diffusion in the passive dumbbell, respectively.} 
{ We also observe that in the passive limit Eq.~(\ref{eq:velocity-cm-average}) reduces to the 
equipartition theorem for the kinetic energy of a point-like particle having  the total mass of the dumbbell  $2m_d$. }
 As, typically, $t_a \gg t_I$, one has 
\begin{equation}
2 m_{\rm d} [\langle {v^2_{\rm cm}}_x\rangle]_{ic} \simeq  k_BT+\frac{m_d F_{\rm act}^2}{\gamma^2}  
\; . 
\label{eq:velocity-cm-average-limit}
\end{equation}

With a similar calculation, starting now from Eq.~(\ref{eq:r-cm}), we calculate
the mean-square displacement (MSD) 
\begin{equation}
\langle \Delta {\mathbf r}_{\rm cm}^2\rangle(t) = [\langle({\mathbf r}_{\rm cm}(t+t_0) - {\mathbf r}_{\rm cm}(t_0))^2\rangle]_{ic}
\label{eq:MSD-def}
\end{equation}
{ with $[ \dots ]_{ic}$ the average over initial conditions at time $t=0$. $t_0$ is a sufficiently long time after preparation such 
that the stationary dynamics have been established and the mean-square displacement is therefore independent of $t_0$.
Henceforth, } $t$ denotes time-delay.

In the limit $t \ll t_I$, 
\begin{eqnarray}
 \langle\triangle{\mathbf r}_{\rm cm}^2\rangle(t)=  2 \langle {v^2_{\rm cm}}_x\rangle  \ t^2
 \; , 
 \label{eq:ballistic1}
\end{eqnarray}
where $\langle {v^2_{\rm cm}}_x\rangle = [\langle {{v^2_{\rm cm}}_x}\rangle]_{ic}$ is the velocity given in 
Eq.~(\ref{eq:velocity-cm-average}). The factor two is due to the sum over the two Cartesian directions. 

In the limit $t \gg t_I$,
\begin{eqnarray}
 &&
 \langle\triangle{\mathbf r}_{\rm cm}^2\rangle (t) =
 4D_{\rm cm}^{\rm pd} \ t 
 \nonumber\\
 && 
 \qquad
 +
 \bigg(\frac{F_{\rm act}}{\gamma}\bigg)^2\frac{2}{D_R}\left(t-\frac{1-e^{-D_Rt}}{D_R}\right)
 \; ,
 \label{msddumbattivasingola}
\end{eqnarray}
where 
\begin{equation}
D_{\rm cm}^{\rm pd}=\frac{k_BT}{2\gamma}
\end{equation} 
is the diffusion constant  in the passive
limit, ${\mathbf F}_{\rm act}=0$, see \cite{Suma14b}.
This equation presents several time scales and limits. (Similar calculations for an active ellipsoid were presented in~\cite{Lowen11}.)
 For $t_I \ll t \ll t_a$ one finds
\begin{equation}
 \langle\triangle{\mathbf r}_{\rm cm}^2\rangle(t)=4D_{\rm cm}^{\rm pd} \ t+
 \bigg(\frac{F_{\rm act}}{\gamma}\bigg)^2t^2 \; ,
 \label{msddumbattivasingola_1limite}
\end{equation}
that can still be split into the passive
diffusive limit $\langle\triangle{\mathbf r}_{cm}^2\rangle=4D_{\rm cm}^{\rm pd} \ t$ for $t_I\ll t<t^*$, and a ballistic
regime  $\langle\triangle{\mathbf r}_{cm}^2\rangle=(F_{\rm act}/\gamma )^2 \ t^2$
for $t^*<t \ll t_a$, where the time scale $t^*$ is given by 
\begin{eqnarray}
t^*&=&\frac{4D_{\rm cm}^{\rm pd}\gamma^2}{F_{\rm act}^2} = \frac{2k_BT \gamma}{F_{\rm act}^2} 
= \left(\frac{4}{\mbox{Pe}}\right)^2 \ \frac{\sigma_{\rm d}^2}{4D_{\rm cm}^{\rm pd}} 
\nonumber\\
&=& 
\left(\frac{4}{\mbox{Pe}}\right)^2 \ t_a
\; . 
\end{eqnarray}
Note that these two intermediate 
regimes {do not exist} if the parameters are such that
$t^*<t_I$ or $t^*>t_a$. They can also be easily confused with super-diffusion $t^\alpha$ 
with $1 < \alpha < 2$ if they are not well separated ($t_I \simeq t^* \simeq t_a$). 
See \cite{Suma14b}  and Fig.~\ref{fig:mean_square_displacement_cm} below for more details.
In the large Pe limit one has $t^* \ll t_a$.
 In the last time-lag regime $t\gg t_a$, we recover normal diffusion,
 \begin{equation}
 \langle\triangle{\mathbf r}_{\rm cm}^2\rangle(t)=4D_A \ t \; ,
 \label{msddumbattivasingola2_limite}
\end{equation} 
with the diffusion coefficient
\begin{equation}
D_A(F_{\rm act}, T, \phi=0) = \frac{k_BT}{2\gamma} \left[ 1+ \frac{1}{2} \left(\frac{F_{\rm act} \sigma_{\rm d}}{k_BT}\right)^2 \right]
\; . 
\label{eq:DA}
\end{equation}
In terms of the diffusion constant of the center of mass of the passive dumbbell and the 
P\'eclet number the above equation  reads $D_A(\mbox{Pe}, \phi=0) = D_{\rm cm}^{\rm pd} \ (1+ \mbox{Pe}^2/8)$.
 
 {In 
 the figures with numerical results 
 for the finite density problem shown in the next Section
 we include data for $\phi=0$ that correspond to the single 
 dumbbell limit.}
 
\section{Finite density systems}
\label{sec:numerical}

In this Section we present {our numerical results}. We focus on three issues: the phase diagram, the 
translational diffusion properties and the rotational diffusion properties.
Details on the numerical method used for solving the dynamical equations (\ref{eqdumbattcoll})
are given in \cite{Suma14b}. We set $m_{\rm d} = \sigma_{\rm d} = k_B = \epsilon = 1$ in proper physical units,
{and} $r_0 = 1.5$, $k =  30$, $\gamma=10$,   assuring over-damped motion and negligible dumbbell vibrations. 
Depending on the plots we used between 15000 and 20000 dumbbells in the simulations.

\subsection{The phase diagram}

Aspects of the phase diagram and the dynamics of this system were already established in~\cite{Suma13,Suma14}. 
It was shown in these papers that at sufficient{ly} low temperature and large active force
 the system phase separates into 
gas-like spatial regions and clusters of agglomerated dumbbells.

\vspace{0.75cm}
\begin{figure*}[ht]
\begin{center}
  \begin{tabular}{cc}
  \includegraphics[scale=0.67]{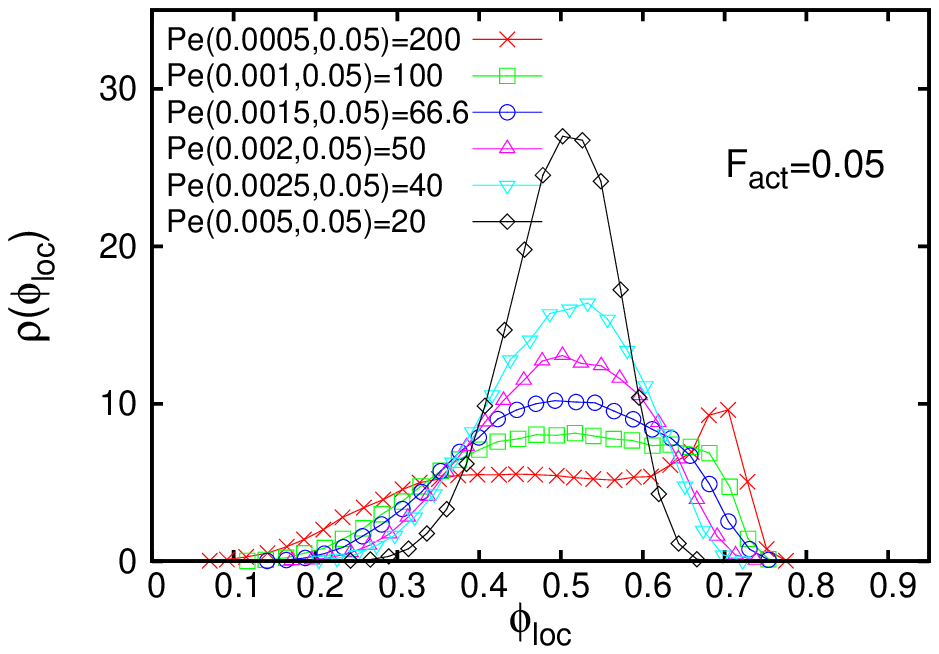}
      \includegraphics[scale=0.67]{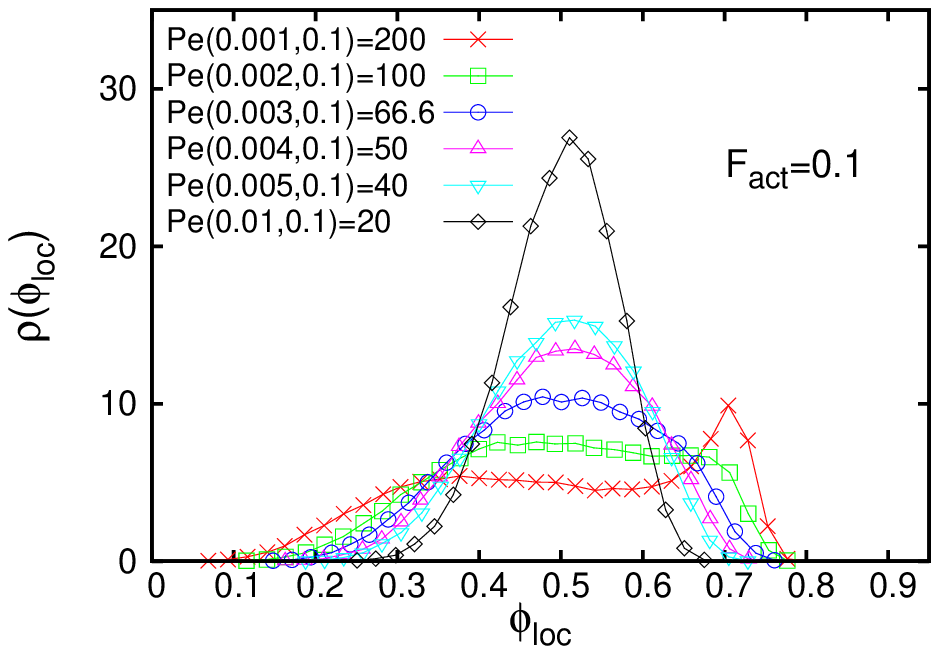}\\
    \includegraphics[scale=0.67]{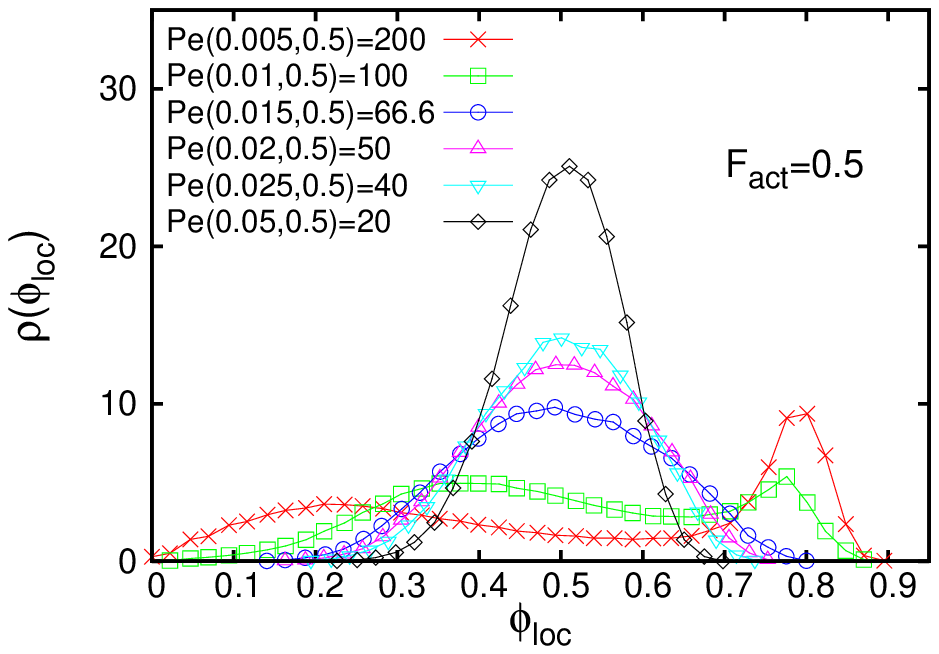} 
     \includegraphics[scale=0.67]{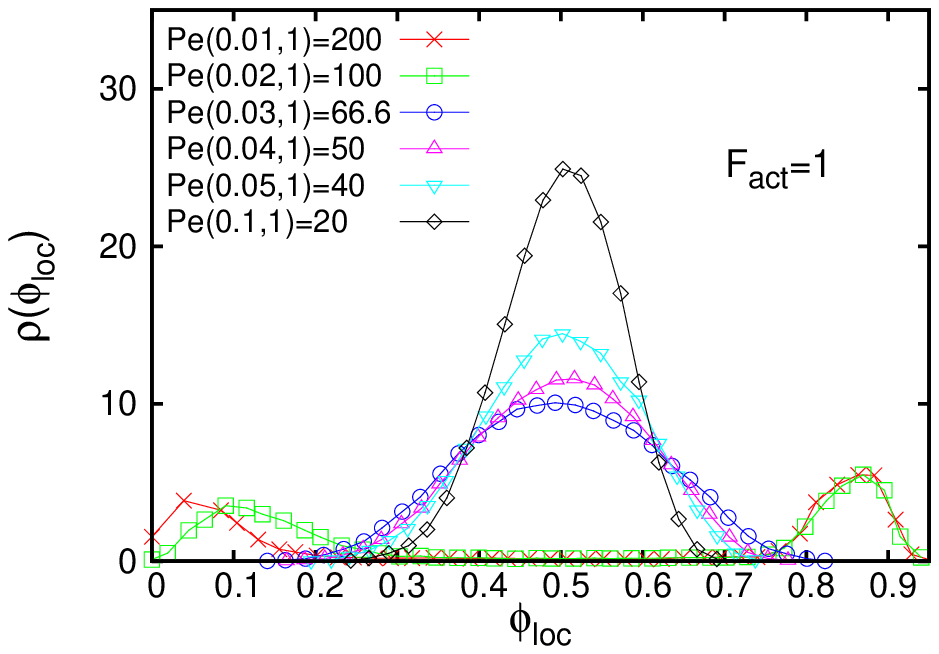}   \\
    \end{tabular}
\caption{ 
Probability distributions for the local density $\phi_{\rm loc}$ at P\'eclet numbers given in the keys in the form Pe$(T,F_{\rm act})$
for different temperatures, and $F_{\rm act}=0.05, \ 0.1, \ 0.5, \ 1$ in the different panels.
The global density of the system is $\phi=0.5$. }
\label{fig:density_distributions} 
\end{center}
\end{figure*}

The model has three important energy scales, $\epsilon$, $k_BT$ and $F_{\rm act} \sigma_{\rm d}$. 
{ There is another energy scale related to the elastic constant $k$
but, since we consider stiff dumbbells in this paper, we do not vary $k$.
Keeping also $\epsilon$ and the  other parameters listed at the beginning of this section fixed},
 we will try to determine whether the phase diagram 
depends upon $k_BT$ and $F_{\rm act} \sigma_{\rm d}$ separately or only though their ratio, {\it i.e.} through the 
P\'eclet number Pe, as usually assumed~\cite{Redner13}. 
The other free parameter to be varied is the global density of the sample, $\phi$.

\begin{figure*}[ht]
\begin{center}
  \begin{tabular}{cc}
	\hspace{-0.9cm}
  \includegraphics[scale=0.75]{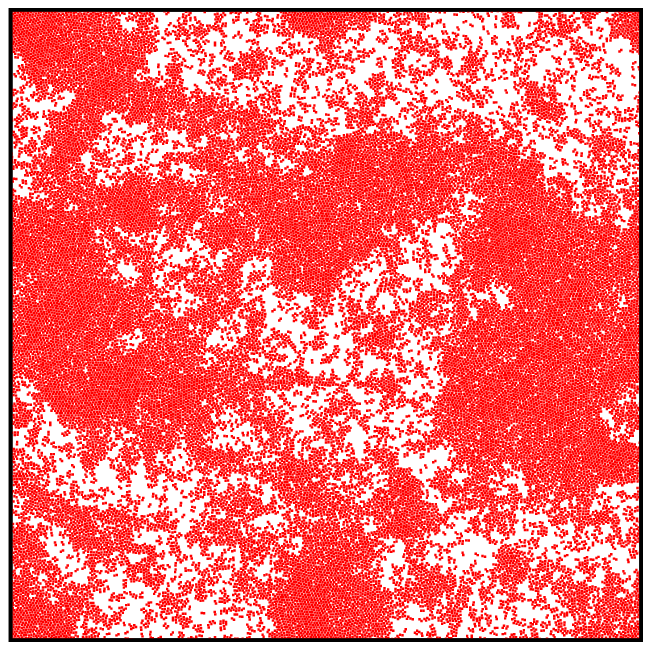}
  \hspace{-0.85cm}
      \includegraphics[scale=0.75]{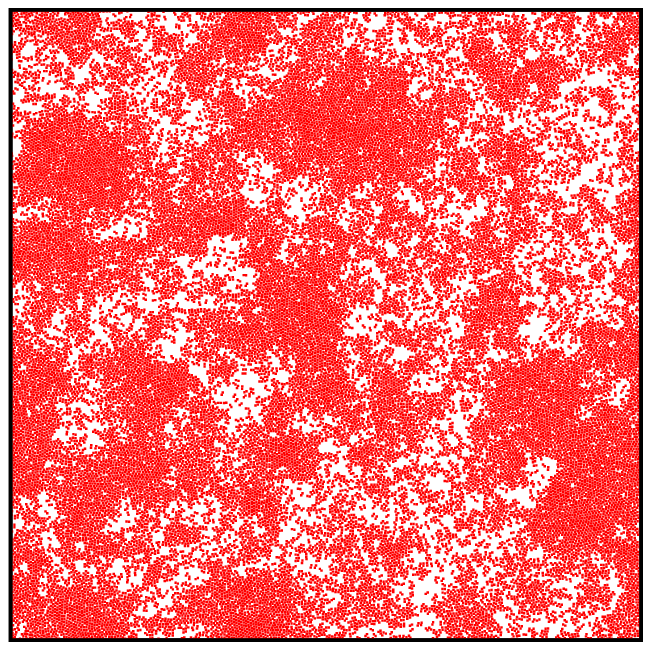}\\
		\hspace{-0.95cm}	
    \includegraphics[scale=0.75]{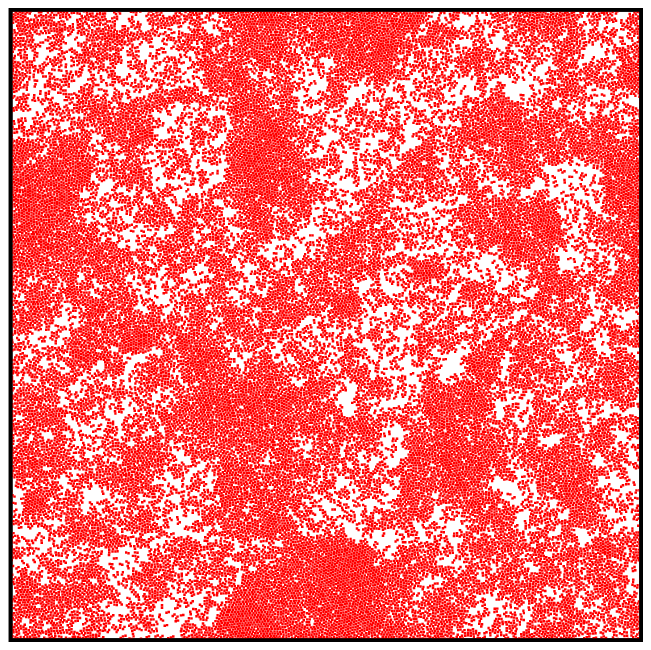} 
  \hspace{-0.85cm}
     \includegraphics[scale=0.75]{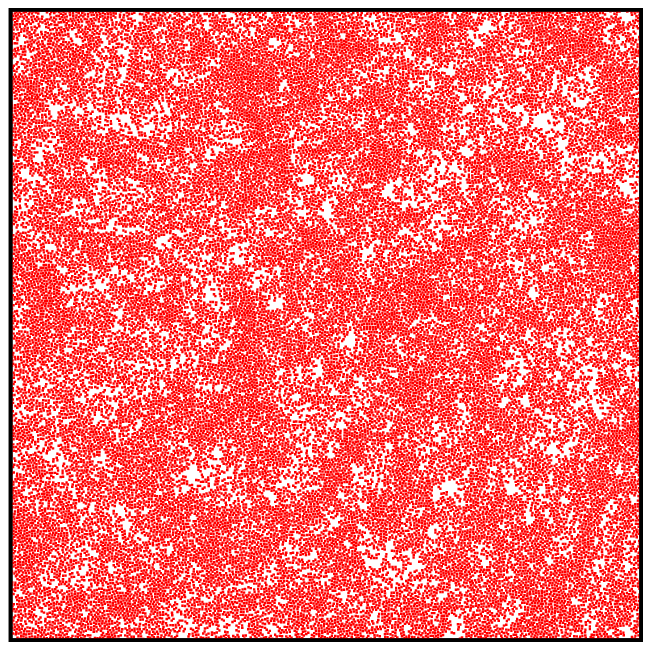}   \\
    \end{tabular}
\caption{ 
Snapshots of the system for  $F_{\rm act}=0.1$ (cfr. the {upper} right panel in Fig.~\ref{fig:density_distributions}),  
and $T=0.001,\ 0.002,\ 0.003 , \ 0.01$ 
corresponding to Pe = $200, \ 100$, both phase separated, Pe = $66$, close to critical, and Pe $= 20$,
homogeneous (from left
to right and from top to bottom). The global density is $\phi=0.5$. 
}
\label{fig:snapshot_phase_transition_Fa0_1} 
\end{center}
\end{figure*}

In the four panels in Fig.~\ref{fig:density_distributions} we show the probability distribution function, $\rho$, 
of the local density, $\phi_{\rm{loc}}$,
for four
values of the active force, $F_{\rm act} = 0.05,  \ 0.1, \ 0.5, \ 1$. 
Each panel contains results for the same set of  P\'eclet numbers 
Pe$(T, F_{\rm act})$ obtained by different combinations 
of temperature and active force.
 The system has global packing fraction $\phi=0.5$.
We used the same operative definition of the local density as in~\cite{Suma13,Suma14}. 
We divided the full system in square plaquettes with linear size $10 \sigma_{\rm d}$ that is 
much smaller than the linear size of the full sample and big enough to sample $\phi_{\rm{loc}}$ correctly.
We improved the statistics by sampling over many different runs of the same kind of system.

At low Pe the system is homogeneous and $\rho$ is peaked around $\phi_{\rm{loc}}=\phi=0.5$.
The critical Pe at which the system starts aggregating is  approximately the same in all panels, 
Pe $ \simeq 65$. Around  this value the density distribution $\rho$  not only becomes asymmetric but starts
developing a second peak at $\phi_{\rm{loc}}> 0.5$
 that characterises the dense phase in the system. Snapshots of typical  configurations 
at $F_{\rm act} = 0.1$ and four values of Pe are shown in Fig.~2. 
 The location of the central peak at  Pe less than the critical value
is independent of all parameters
(apart from $\phi$) while the location of the  peak at $\phi_{\rm{loc}}>0.5$ is situated at  different values of $\phi_{\rm{loc}}$ for
different $F_{\rm act}$ and the same Pe (compare the different panels in Fig.~\ref{fig:density_distributions}). The 
reason for this is that the strength of the interactions between the dumbbells under different $F_{\rm act}$ 
is different as $F_{\rm act}/\epsilon$ varies with $F_{\rm act}$. 
A larger active force permits the dumbbells to {be more compact}, while a lower one favors looser clusters. 

{
We repeated the analysis above for the cases with total packing fractions $\phi=0.3, \ 0.4, \ 0.6$.
We found the same values 
for the densities of the separated phases  at $\phi=0.4, \ 0.6$ and  Pe = $100, \ 200$, and at $\phi=0.3$ and Pe = 200.
At $\phi=0.3$ and $\rm{Pe} \le 100$  the effects of the presence of the  spinodal  line require a more elaborate analysis 
of the phase diagram, as discussed in \cite{Suma14}.}  
In Table~\ref{table1}  we report these density values
for  the cases with $F_{\rm act} = 1, \ 0.5, \ 0.1, \ 0.05$ and Pe $=100, \ 200$. 
As observed, the coexistence values get closer for smaller active forces even though the 
P\'eclet number remains the same.

\vspace{0.75cm} 

\begin{table}[h]
\centering
\begin{tabular}{|l|c|c|c|c|}
\hline  
 \ Pe \ & $\ F_{\rm act}=0.05 \ $  &	$\ F_{\rm act}=0.1 \ $   &	$\ F_{\rm act}=0.5 \ $  &	$\ F_{\rm act}=1 \ $ \\
 \hline
\ 200  \ & 0.37	 & 0.34       &	0.21    &	0.049 \\
\                & 0.70	&  0.71     &	0.80    & 	0.890 \\
\ 100   \ & 0.44 	& 0.41	     &	0.37     &	0.096 \\
\                & 0.66      & 0.68	     &	0.77	  &	0.870 \\
\hline
\end{tabular}
\caption{Density values of the two coexisting phases measured from  the histograms in Fig.~\ref{fig:density_distributions}
{ at $\phi=0.5$. The first and the third lines refer to the dilute phase while the two other lines correspond to the aggregated phase. Similar 
values are obtained at $\phi=0.4,0.6$ (Pe = $100, \ 200$) and at $\phi=0.3$
(Pe $= 200$).
}
}
\label{table1}
\end{table}

{In Fig.~\ref{fig:snapshot_phase_transition_Fa0_1} we show four snapshots of the 
system configuration. The active force is $F_{\rm act}=0.1$ in all panels and temperature is increased
from left to right and from top to bottom. The configuration in the upper-left panel (Pe = 200) shows phase separation with 
large scale clusters while the configuration in the lower-right panel (Pe = 20) is clearly homogeneous. The case 
Pe = 100 is in the segregated phase while the one for Pe = 66 is close to critical.
}


\subsection{Translational diffusion properties} 

In~Ref.~\cite{wu} the diffusion properties of a tracer immersed in a bacterial bath were monitored. 
A cross-over between a super-diffusive regime at short time-delays and a diffusive regime 
at long time-delays was reported. The cross-over time was found to increase linearly with the density of the 
active medium, showing that the cross-over is not due to the tracer's inertia but to the dynamical properties of the 
bacterial bath. We explore here the same issues by focusing on the MSD
of the center of mass of the dumbbells, defined in Eq.~(\ref{eq:MSD-def}). We will consider, for the rest of the paper, 
sufficiently low P\'eclet numbers  such  that the system 
will  always be in the  homogenous phase even though fluctuation effects can be relevant, as we will see.

{
\subsubsection{Dumbbell trajectories}
}

\begin{figure*}[th]
\begin{center}
  \begin{tabular}{cc}
      \includegraphics[scale=0.67]{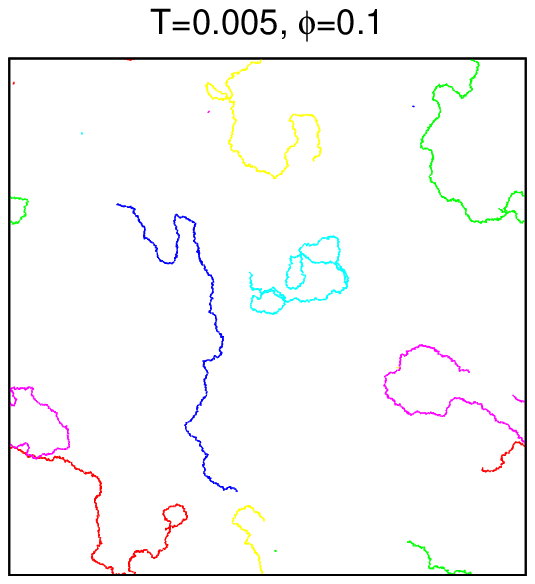}
      \hspace{-1.5cm}
     \includegraphics[scale=0.67]{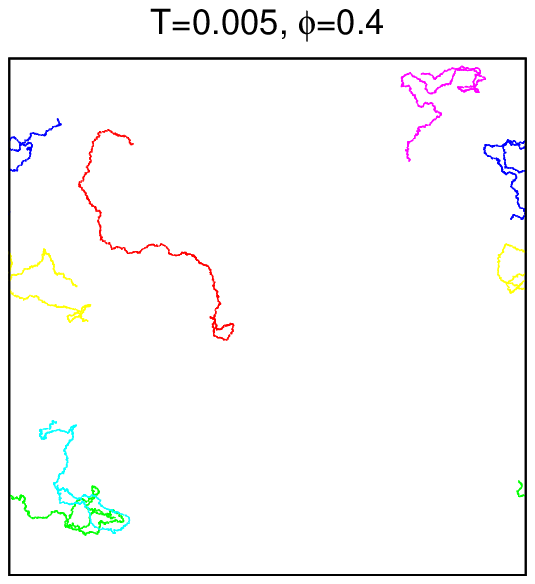}    \\
     \includegraphics[scale=0.67]{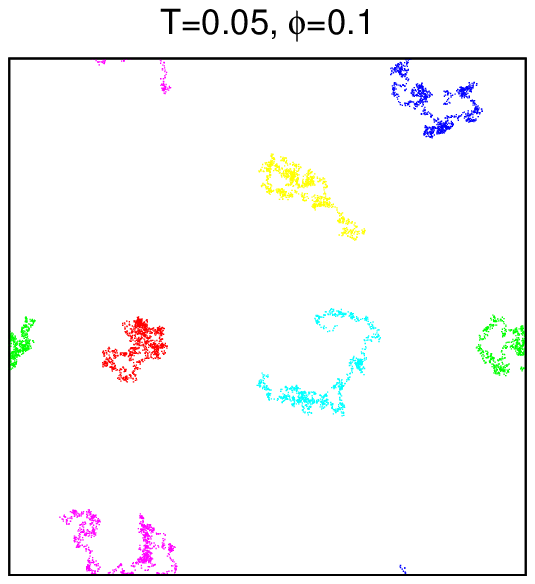}    
           \hspace{-1.5cm}
     \includegraphics[scale=0.67]{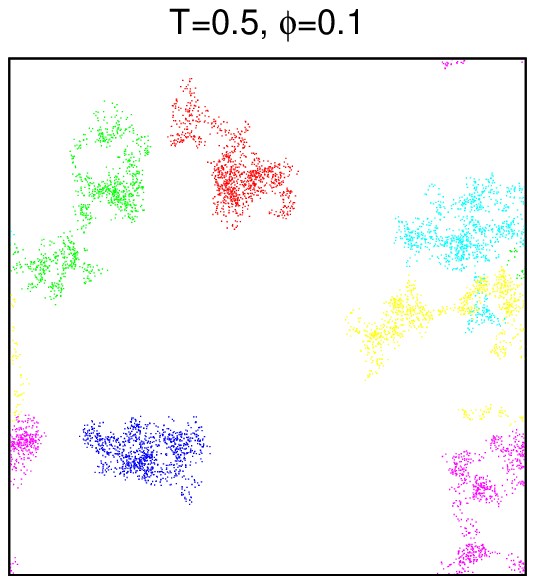}    \\
    \end{tabular}
\caption{Several trajectories of one bead in different dumbbells. The active force is $F_{\rm act}=0.1$ 
and the evolution time interval is $t=10000$ in 
all panels. The other parameters
are $T=0.005$  and $\phi=0.1,\ 0.4$ (upper panels, left and right), and $T=0.05 ,\ 0.5$ and
$\phi=0.1$ (lower panels, left and right).
{ 
The values of the single dumbbell diffusion coefficient, as calculated from Eq.~(\ref{eq:DA}), are  
$D_A = 0.050,\ 0.0075,\ 0.025$ for the cases at $T=0.005,\ 0.05, \ 0.5$, respectively.}
}
\label{fig:spostamento} 
\end{center}
\end{figure*}

Several single dumbbell trajectories are shown in~Fig.~\ref{fig:spostamento} for different values of the
temperature and global density, under the same active force $F_{\rm act} = 0.1$.
The trajectories correspond  to  a total time  interval that includes the late  diffusive regime (see below).
At low temperature and global density ($T=0.005$ and $\phi=0.1$, upper left panel) we see periods of 
long directional motion. These are 
reduced at higher global density ($\phi=0.4$ upper right panel). Increasing temperature { at $\phi=0.1$
 ($T=0.05$ and $T=0.5$ 
lower left and right panels, respectively) the trajectories become more similar to the typical ones of passive 
diffusion. While the trajectories are very stretched at $T=0.005$, they  become  the most compact in the intermediate case
at $T=0.05$ and again quite stretched in the last case at $T=0.5$.
This  behaviour  corresponds to the non monotonic behavior of 
 the translational diffusion constant of  Eq.~(\ref{eq:DA})  in terms of temperature.
 It  decreases going from  $T=0.005$ to $T=0.05$ while it  increases going from $T=0.05$ to  $T=0.5$.
 The single dumbbell diffusion coefficient, as calculated from Eq.~(\ref{eq:DA}),is  
$D_A = 0.050,\ 0.0075,\ 0.025$ for the cases at $T=0.005, \ 0.05, \ 0.5$, respectively.
}

\vspace{0.75cm}

{
\subsubsection{Four dynamic regimes}
}

\begin{figure*}[ht]
\begin{center}
  \begin{tabular}{cc}
      \includegraphics[scale=0.67]{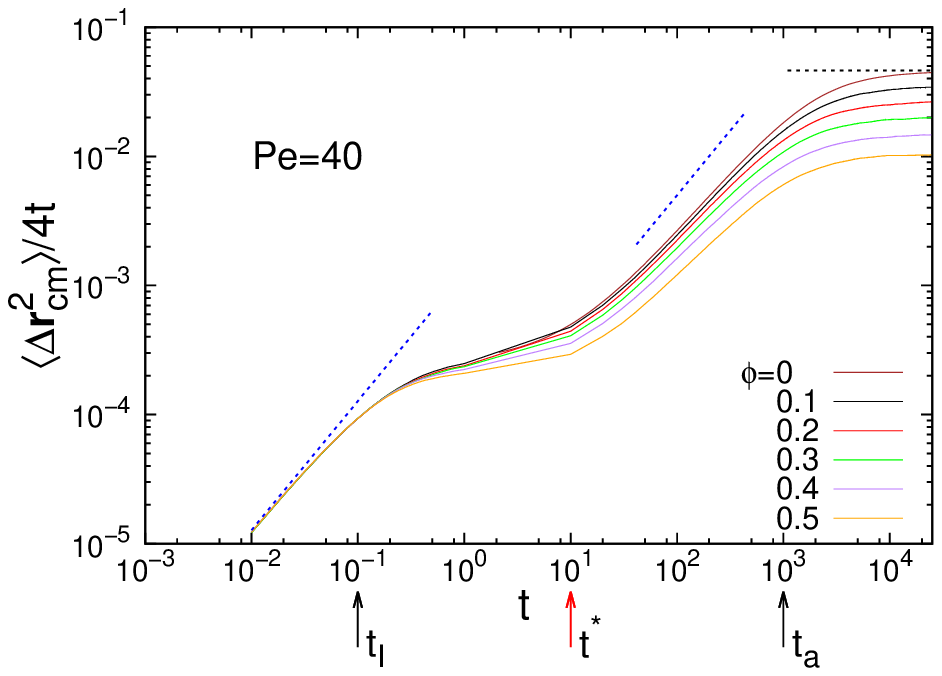}
     \includegraphics[scale=0.67]{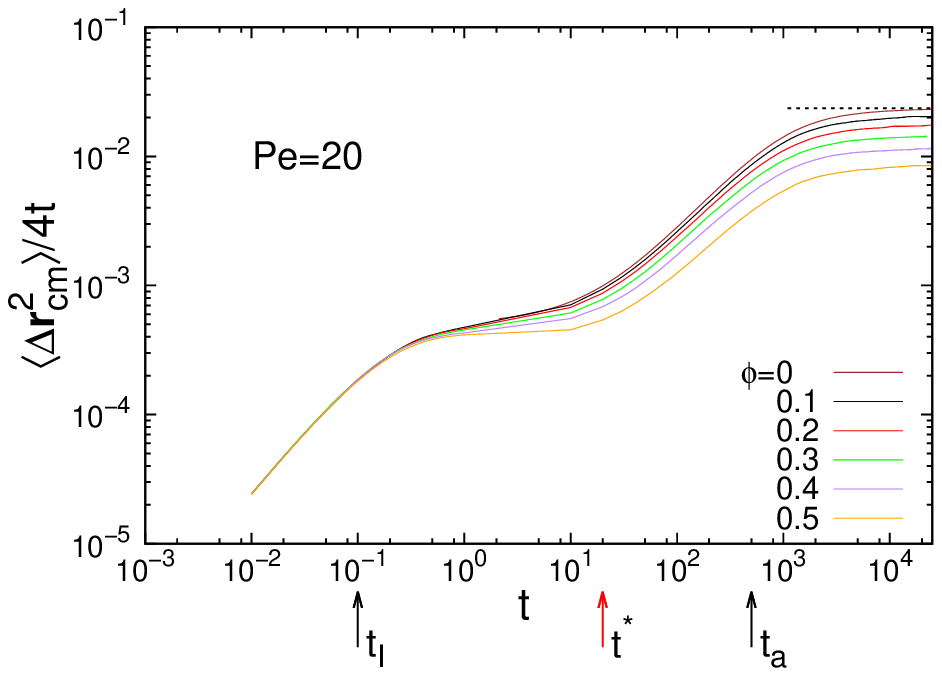}    \\
    \includegraphics[scale=0.67]{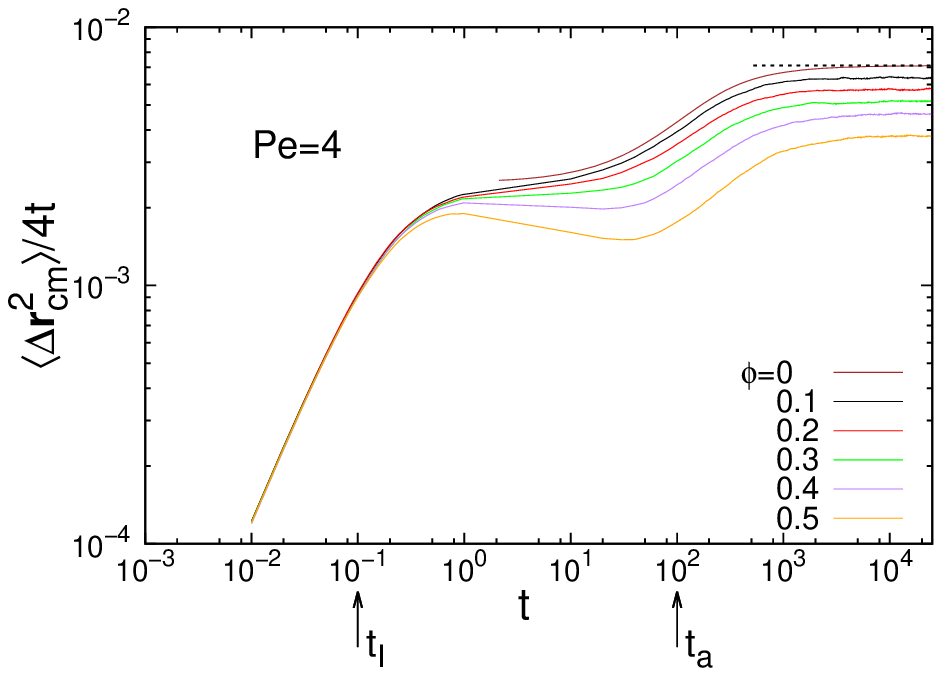}
          \includegraphics[scale=0.67]{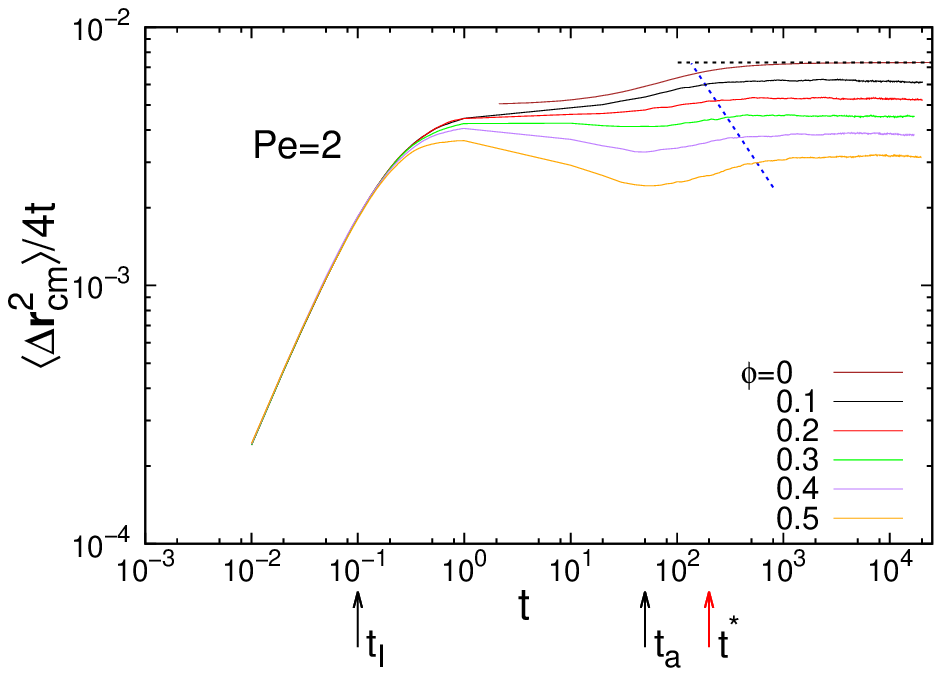}    
    \end{tabular}
\caption{The center of mass MSD normalised by time-delay, for an active system 
under $F_{\rm act}=0.1$ at $T=0.005, \ 0.01,\ 0.05,\ 0.1$ (Pe = $40, \  20, \ 4, \ 2$), 
at the different densities given in the key.
The P\'eclet number decreases from one panel to the other. 
It  induces a strong qualitative change in $\langle \Delta {\mathbf r}_{\rm cm}^2 \rangle$, 
see the text for a detailed discussion. The two dashes in the first panel represent the ballistic dependence $\simeq t^2$. 
The dashed line in the last panel is a guide-to-the-eye for the density dependence of the last cross-over time-delay
that increases weakly with $\phi$. { The horizontal dotted lines at long times correspond to the values 
of the single dumbbell diffusion constant $D_A$ from Eq.~(\ref{eq:DA}).}
The vertical black arrows indicate the single dumbbell time-scales $t_I$ and $t_a$, while the  red arrows
indicate the single dumbbell characteristic time $t^*$, for each case. In the case with $\mbox{Pe} = 4$ 
the times $t^*$ and $t_a$ coincide. {
 The curves for $\phi=0$ are obtained from Eq.~(\ref{msddumbattivasingola}), 
valid in the limit $t \gg t_I=0.1$; for this reason they start  from the middle of the graph.}
}
\label{fig:mean_square_displacement_cm} 
\end{center}
\end{figure*}

In Fig.~\ref{fig:mean_square_displacement_cm} we show  the center of mass MSD 
normalised by time-delay in such a way that normal diffusion appears as a plateau. The four panels 
display data at four temperatures,
$T=0.005, \ 0.01,$ $0.05, \ 0.1$, all under the same active force $F_{\rm act} =0.1$. Each panel has five 
curves in it, corresponding to five different densities given in the key. In all cases $m_{\rm d}=1$ and $\gamma=10$ implying $t_I=0.1$. 
The characteristic times $t_I, \ t^*, \ t_a$ are shown with small vertical arrows in each panel. 
These plots show several interesting 
features:\\
-- In all cases there is a first ballistic regime (the dashed segment close to the data is a guide-to-the-eye) 
with a pre-factor that is independent of $\phi$ and increases with temperature as given by Eq.~(\ref{eq:ballistic1}) 
(The case $t\ll t_I$ of the single  dumbbell.)
\\
-- Next,  the dynamics slow down and, depending on $T$ and $\phi$, the normalised mean-square 
displacement may attain a plateau associated to normal diffusion (low $T$) or even decrease, suggesting sub-diffusion.
(The case $t_I \ll t \ll t^* \ll t_a$ of the single dumbbell.)
\\
-- The dynamics accelerate next, with a second super-diffusive regime in which the curves 
for all $\phi$ in each panel look approximately parallel and very close to ballistic at $T=0.005, \ 0.01, \ 0.05$. 
(The case $t_I \ll t^* \ll t \ll t_a$ of the single dumbbell.) 
\\
-- Finally, the late normal diffusive regime is reached with all curves saturating at a plateau that yields the 
different $D_A$ coefficients.
(The case $t_I \ll t^*, \ t_a \ll t$  of the single dumbbell.)

It is hard to assert whether  the intermediate regime is super-diffusive or simply ballistic as the time-scales 
$t^*$ and $t_a$ are not sufficiently well separated (and not even ordered as $t^*< t_a$ in the last panel). 
Moreover, in the last two panels (high $T$ or low Pe) the diffusion-ballistic-diffusion regimes are mixed,
due to the fact that the condition $t^* \ll t_a$ is no longer satisfied.  
The  effective slope in the intermediate  super-diffusive regime
 decreases when the density increases. 
 
\begin{figure*}[ht]
\begin{center}
\includegraphics[scale=0.67]{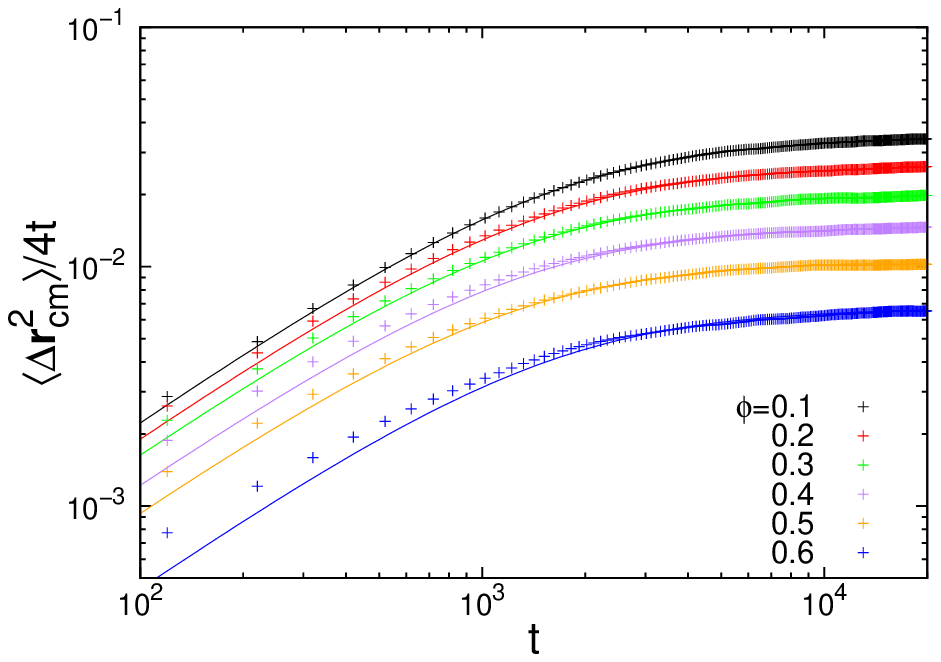}
\includegraphics[scale=0.67]{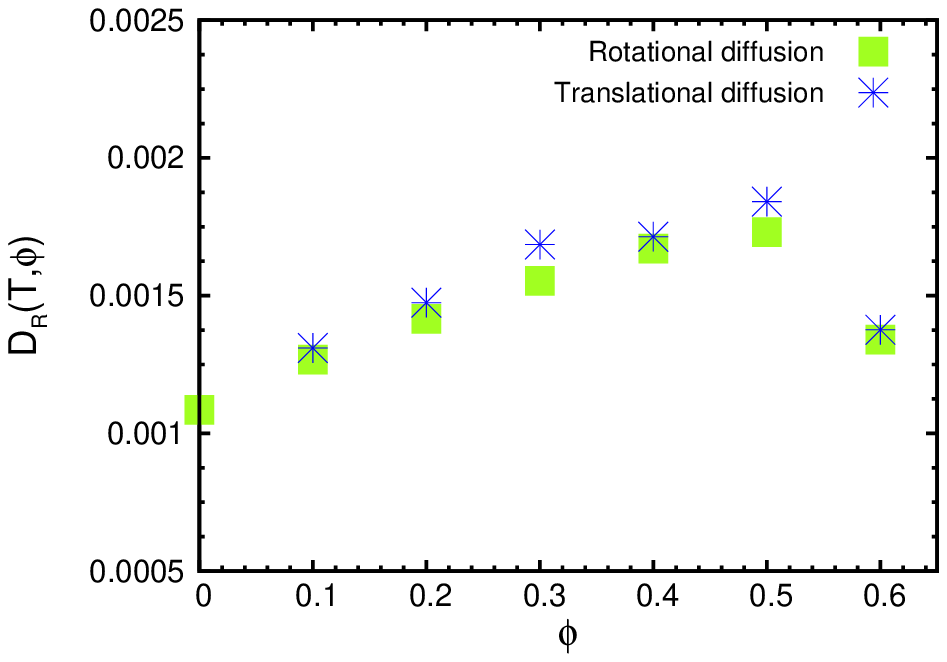}
\caption{Left panel: Fit of the center of mass MSD normalised by time-delay with the expression in 	
Eq.~(\ref{fitmsddumbattivasingola}). Pe $=40$ and the different data sets correspond to the global densities given in the key. 
Right panel: { The rotation diffusion coefficient as extracted from the fit  of the center--of--mass MSD shown 
in the left panel 
($D_R^{\phi}$ in Eq.~(\ref{fitmsddumbattivasingola})  - blue stars), 
and 
from the late-time diffusive rotational MSD directly measured as shown in  Sec.~\ref{sec:rotational} (green squares).
} { Statistical errors are within the  size of the symbols used in the figure.}
}
\label{fig:mean_square_displacement_cm-fit}  
\end{center}
\end{figure*}

 A rather good fit of the finite density data in the limit Pe $\gg 1$ and for time-delays such that $t \ge t^{*}$
 is achieved by using the single dumbbell  expression in Eq.~(\ref{msddumbattivasingola}) 
  \begin{equation}
 \langle\triangle{\mathbf r}_{\rm cm}^2\rangle (t)=
 4D_A^{\phi}\bigg(t-\frac{1-e^{-D_R^{\phi}t}}{D_R^{\phi}}\bigg)
 \; ,
 \label{fitmsddumbattivasingola}
\end{equation}
without the first term (negligible if Pe $\gg 1$) and upgrading the remaining parameters, $D_A^{\phi}$ and $D_R^{\phi}$, 
to be density-dependent fitting parameters, as  done in~\cite{wu,Fily12}. This is shown in
Fig.~\ref{fig:mean_square_displacement_cm-fit} (left panel). For not that large values of Pe
one could recover the remaining parameter and use instead
$\langle\triangle{\mathbf r}_{\rm cm}^2\rangle (t)=
 4D^{{\rm pd}, \phi}_{\rm cm} t +4D_A^{\phi} (t-\frac{1-e^{-D_R^{\phi}t}}{D_R^{\phi}} )$
 with an additional fitting parameter. 
Figure~\ref{fig:mean_square_displacement_cm-fit} (right panel)
also shows a good agreement between the values of $D_R^{\phi}$ found in these fits
and the values of the rotation diffusion coefficient $D_R(F_{\rm act},T,\phi)$ 
coming from  the late time-delay diffusive regime in the rotational MSD discussed in Sec.~\ref{sec:rotational}.  
 
The cross-over time-delay between the last ballistic or super-diffusive, and the diffusive regimes seems quite $\phi$-independent
in the first two panels $T=0.005, \ 0.01$ and it increases, though rather weakly, with $\phi$, 
in the last two panels, $T=0.05, \ 0.1$, see the inclined dashed line in the last panel that is also a guide-to-the-eye. This 
cross-over time-delay is the one that we could associate to 
{ 
the cross-over time between a superdiffusive regime and the last diffusive regime found in the experiment
}
in~\cite{wu}.
The strongest effect of density is though on the first diffusive or sub-diffusive regime.

{In summary, no large qualitative change in the center of mass MSD
behavior is observed in the range $\phi \in [0, 0.6]$. There is just a natural slowing down of the dynamics 
with larger packing fractions that translates into a change from diffusive to sub-diffusive behavior in the 
second regime and a general decrease of the diffusion constant in the last regime for all Pe. We study the
dependence of the diffusion constant with Pe in 
detail below.}

\vskip 1cm

\subsubsection{The late-epochs translation diffusion coefficient}

Let us now discuss the normal diffusive regime at longest time-lags.
{
In~\cite{Suma14b} we studied 
 the translational diffusion coefficient
$D_A$ as a function of $F_{\rm act}$ and 
$\phi$ at fixed temperature. In particular, we compared the $\phi$ dependence 
to the Tokuyama-Oppenheim law for colloids~\cite{*[{}] [{. In this work  the diffusion coefficient for a colloidal system at finite density $\phi$ is evaluated as 
	$D(\phi)/D(0) = (1+H(\phi))^{-1}$, with $H(\phi)$ a  function of $\phi$ 
	without free parameters  reducing to  a linear decreasing of D($\phi$) at small $\phi$. }] Tokuyama}. 
Here, we first examine, instead, the $T$ and $\phi$ dependence of $D_A$ 
for fixed active force, $F_{\rm act}$. Then we consider how the dependence  of  $D_A$ from $T$ and $F_{\rm act}$
can be re-expressed in terms of 
the P\'eclet number.
The main results for $D_A$ obtained in~\cite{Suma14b} will also be revisited  in this subsection.
}

The first question we want to answer is whether $D_A$ depends on $k_BT$ as for the 
single dumbbell case ($\phi=0$), the functional form recalled in Eq.~(\ref{eq:DA}). For $F_{\rm act} \sigma_{\rm d}/k_BT \ll 1$
such that the quadratic term can be neglected this equation implies the  linear growth of $D_A$ with $k_B T$
{as in the passive limit}. 
Instead, when the second term dominates, {\it i.e.} for very small thermal energy with respect
to the work performed by the active force, $D_A$ should decay as 
$1/(k_BT)$ with a slope that is quadratic in $F_{\rm act} \sigma_{\rm d}$.  

\begin{figure*}[ht]
\begin{center}
  \begin{tabular}{cc}
      \includegraphics[scale=0.9]{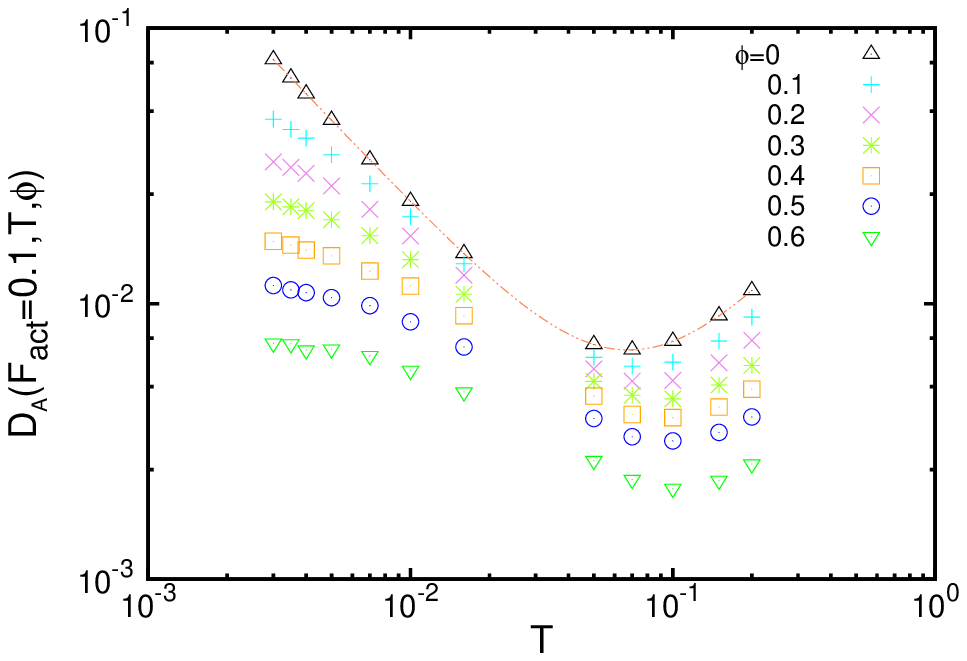}
      \\
    \end{tabular}
\caption{The center of mass diffusion coefficient, $D_A$, as a function of temperature, $T$, at 
different densities $\phi$ given in the key and fixed active force, $F_{\rm act}=0.1$. 
The triangular data points joined by a dotted curve labeled $\phi=0$ are given by the 
theoretical $D_A$ with $\sigma_{\rm d}=0.96$.
}
\label{fig:diffusion_cm_Fa0_1_varying_T} 
\end{center}
\end{figure*}

In Fig.~\ref{fig:diffusion_cm_Fa0_1_varying_T} we
display $D_A$ as a function of $T$ for various values of $\phi$ given in the key and $F_{\rm act}=0.1$. 
The theoretical values for $\phi=0$ are included in the figure (with open triangles joined by a dotted curve).
Here, we used  the measured value for the distance between the centres of the
two colloids, that is  $r \approx 0.96 \ \sigma_{\rm d}$. {In the rest of this section 
we simply call $\sigma_{\rm d}$ the molecular length and we take $\sigma_{\rm d} = 0.96$.}

\vspace{0.75cm}

The error-bars are smaller than the symbol size and we do not display them.
The curves show a minimum located at 
$k_BT_{\rm min}=F_{\rm act}\sigma_{\rm d}/\sqrt{2}$ for $\phi = 0$, that weakly increases with $\phi$.
The two regimes, Pe $\ll 1$ and Pe $\gg 1$, still 
exist and $D_A$  is dominated by thermal fluctuations in the former and 
by the work done by the active force in the latter as in the single dumbbell limit. 
We see a saturation of $D_A$ at small values of $T$ for $\phi >0.2$ and therefore the breakdown of the 
single dumbbell $1/(k_BT)$ behaviour at low temperatures. Instead, at high temperatures 
$D_A$ seems to retain the linear growth with temperature of the single dumbbell at least for the temperatures
used in the simulations.

\begin{figure*}[ht]
\begin{center}
  \begin{tabular}{cc}
      \includegraphics[scale=0.67]{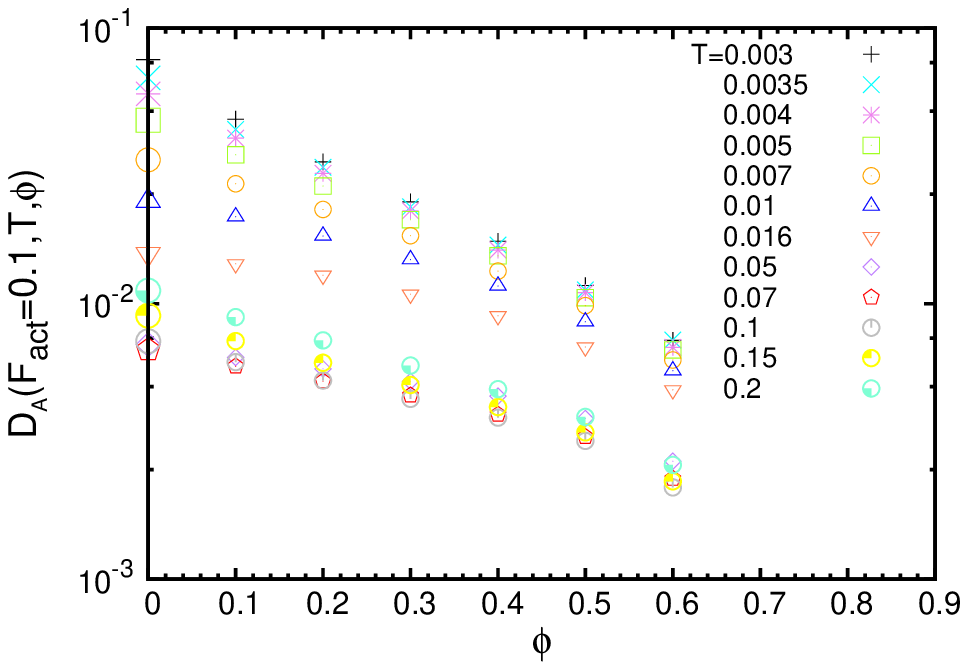}
    \includegraphics[scale=0.67]{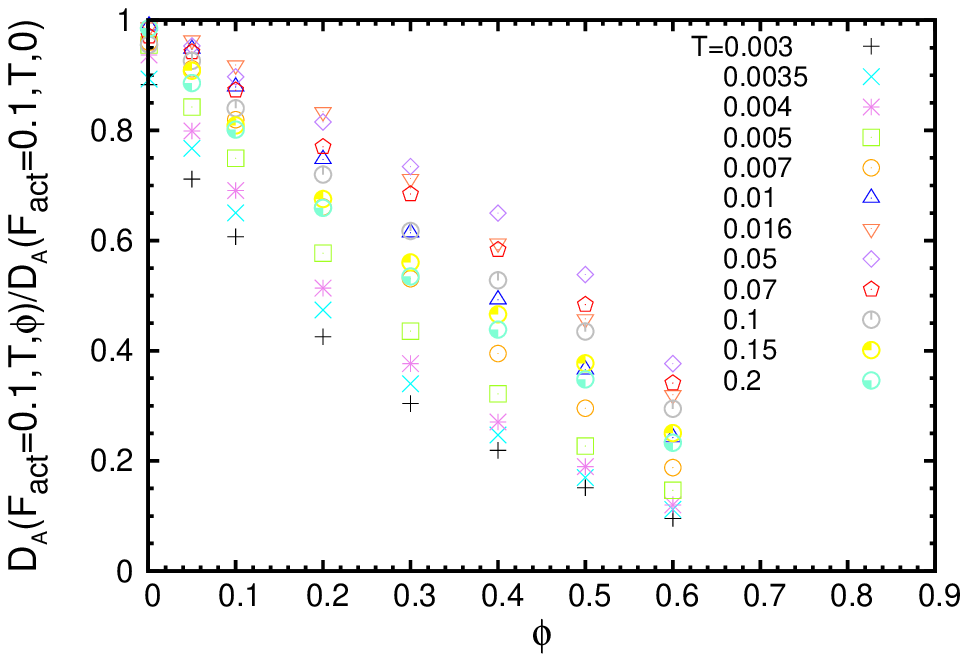}\\  
    \end{tabular}
\caption{Center of mass diffusion constant at various $T$ given in the key (left panel) 
and center of mass diffusion constant over the one for one dumbbell (right panel) both as a function of 
$\phi$. {  The smallest densities considered are $\phi=0$ in the left panel and $\phi=0.001$ in the right panel.  } The active force is fixed to $F_{\rm act}=0.1$. The symbol convention is the same in the two panels.
}
\label{fig:diffusion_cm_Fa0_1} 
\end{center}
\end{figure*}

Figure~\ref{fig:diffusion_cm_Fa0_1_varying_T} also shows that for the Pe that we used
$D_A$ is a decreasing function of  $\phi$ at all fixed temperatures. This fact can be better appreciated in the left  
panel in Fig.~\ref{fig:diffusion_cm_Fa0_1}, where $D_A$ is plotted as a function of $\phi$ for various
temperatures given in the key. (Recall that the $\phi$ dependence of $D_A$ at fixed $T$ and 
for different active forces was discussed in~\cite{Suma14b} where it was shown how 
the Tokuyama-Oppenheim~\cite{Tokuyama} law of the passive system was simplified under activation to a
decay that is  close to a simple exponential. We will come back to this issue below.)

The non-monotonicity of $D_A$ as a function 
of $T$ already discussed in Fig.~\ref{fig:diffusion_cm_Fa0_1_varying_T} 
 is confirmed by the data presentation in Fig.~\ref{fig:diffusion_cm_Fa0_1}, with the minimum 
 situated around $T \simeq 0.07$. In the right panel we observe
the opposite behaviour in the ratio $D_A(F_{\rm act}=0.1,T,\phi)/D_A(F_{\rm act}=0.1,T,0)$, 
 first growing for increasing $T$ to reverse its trend at around
$T\simeq 0.05 - 0.07$. Consistently with the  behaviour 
found in~\cite{Suma14b},  there are  temperatures such that the data for the above ratio cross 
each other when the density is increased, see for example  
$T=0.01, \ 0.1$ (or Pe $=2, \ 20$).  { The right panel in Fig.~\ref{fig:diffusion_cm_Fa0_1}
also shows that a very small density can have relevant effects on the behaviour of the diffusion coefficient.}  

We have repeated this analysis for a stronger active force and we found that the results are 
consistent, with a cross-over temperature that grows with $F_{\rm act}\sigma_{\rm d}$,
as predicted by the single dumbbell equation, though we cannot assert that the dependence
be linear. 

\vspace{0.75cm}

\begin{figure*}[ht]
\begin{center}
  \begin{tabular}{cc}
      \includegraphics[scale=0.67]{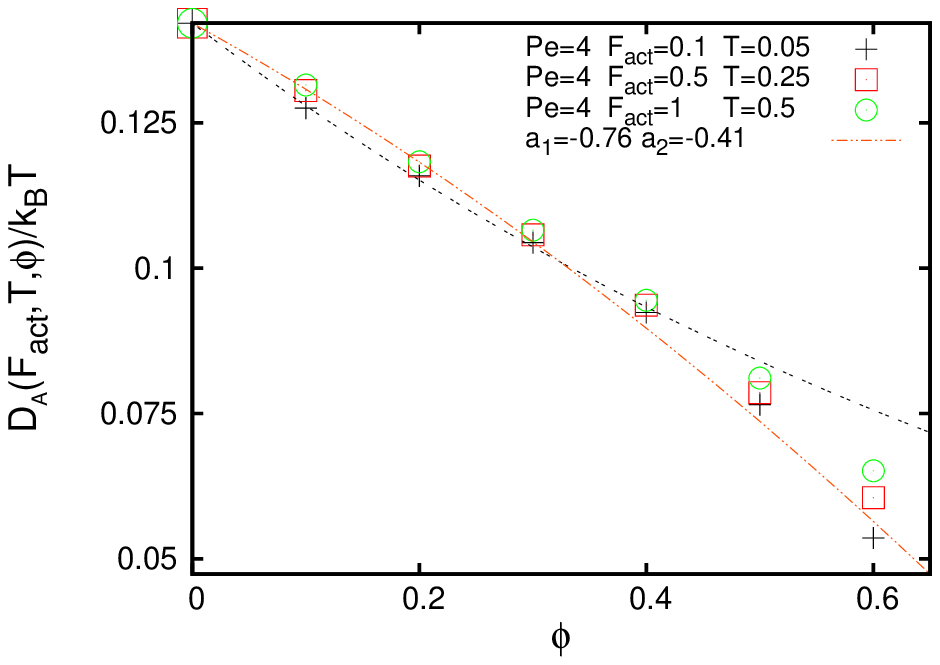}
    \includegraphics[scale=0.67]{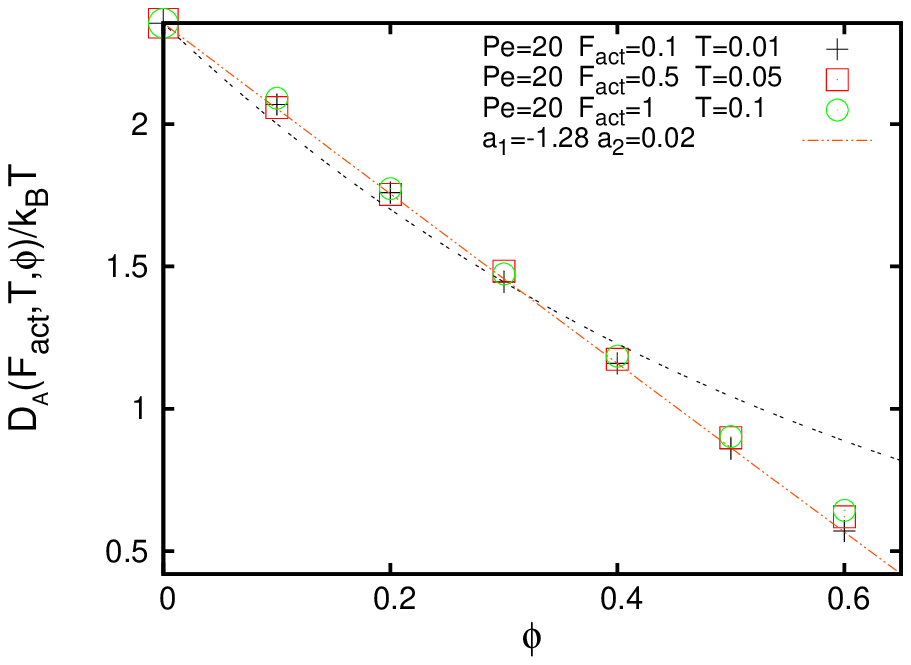}  
    \\
    \includegraphics[scale=0.67]{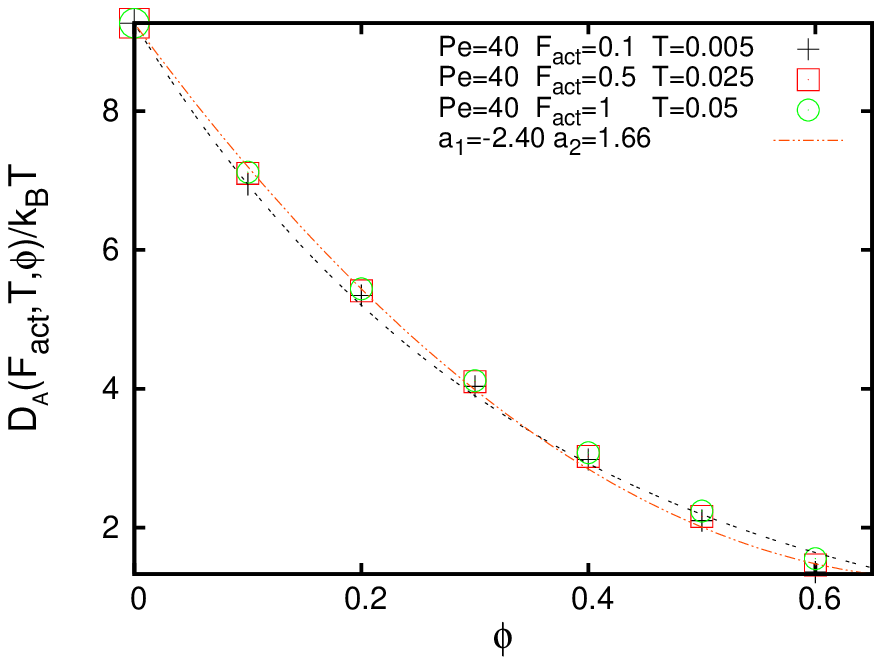}
    \includegraphics[scale=0.67]{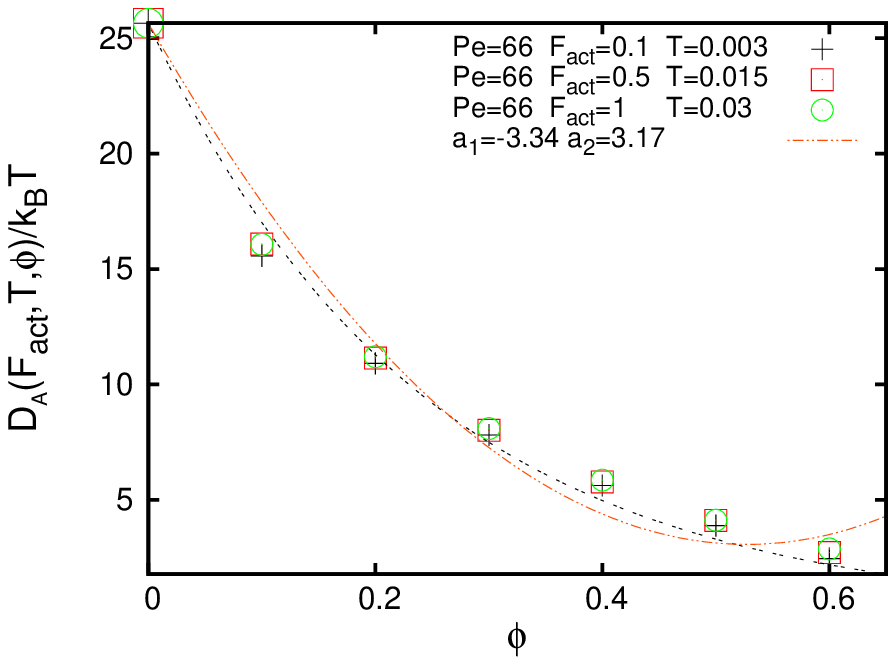}\\
    \end{tabular}
\caption{Center of mass diffusion constant, $D_A$, 
over $k_BT$ for Pe $= 4, \ 20,\ 40, \ 66$ as shown in the keys. For each value of the 
P\'eclet number three couples of values of $F_{\rm act}$ and $T$ were used. The black dotted  lines are exponential fits
to the data points, as suggested by Eq.~(\ref{eq:exponential-fit}). The red line-points are the quadratic fit in Eq.~(\ref{eq:quadratic-fit})
{with fitting parameters given in the keys}. 
See the text for a discussion. 
}
\label{fig:diffusion_cm} 
\end{center}
\end{figure*}

Next, we  analyse in Fig.~\ref{fig:diffusion_cm}  whether the ratio of diffusion coefficients
of the active system at finite density and single passive dumbbell $D_A(F_{\rm act},T,\phi)/D_{\rm cm}^{\rm pd}$
depends only on the P\'eclet number, as it  does for the single dumbbell problem. With this aim, we  fix $F_{\rm act}$ and we vary $T$, 
and the values Pe = $4, \ 20, \ 40, \ 66$ in each panel are obtained from three different combinations of $F_{\rm act}$ and $T$. 
 In all panels the collapse of data is very good. Note the change in concavity of the 
 collapsed data  that occurs at Pe = 20. This value is relatively far from the transition between homogeneous and segregated phases
 estimated in~\cite{Suma14,Suma14b}, and the system configurations are still homogeneous, see
the last panel in Fig.~\ref{fig:snapshot_phase_transition_Fa0_1}, though with a distribution of local densities, $\phi_{\rm{loc}}$, 
with a certain width, see Fig.~\ref{fig:density_distributions}.
 
 These results suggest
 \begin{equation}
 D_A(F_{\rm act}, T, \phi) = k_BT \  f_A\left( \mbox{Pe}, \phi\right)
  \end{equation}
 with $f_A(\mbox{Pe}, 0) = (2\gamma)^{-1} (1 + \mbox{Pe}^2/8) = D_A(F_{\rm act}, T, 0)/(k_BT)$ and 
 $f_A$ a decreasing non-linear function of $\phi$ at fixed Pe.
This relation is equivalent to
\begin{equation}
 \frac{D_A(F_{\rm act}, T, \phi)}{D_A(F_{\rm act}, T,0)} = \frac{f_A(\mbox{Pe}, \phi)}{f_A(\mbox{Pe}, 0)}
 \; . 
\end{equation}

\vspace{0.5cm}

The l.h.s. is what we studied in~\cite{Suma14b} as a function of $F_{\rm act}$ and $\phi$, keeping $T$ fixed, and we proposed
\begin{equation}
 \frac{D_A(F_{\rm act}, T, \phi)}{D_A(F_{\rm act}, T,0)} = e^{-b(F_{\rm act}) \phi}
\end{equation}
with $b$ { a non-monotonic fitting function of} $F_{\rm act}$.
Knowing now that $D_A/(k_BT)$ depends on $F_{\rm act}$ and $T$ only through  Pe, we deduce 
\begin{equation}
D_A(F_{\rm act}, T, \phi) = D_{\rm cm}^{\rm pd} \ (1+\mbox{Pe}^2/8) \ e^{-b({\rm Pe}) \phi} \; . 
\label{eq:exponential-fit}
 \end{equation}
 Note that in~\cite{Suma14b} the maximum in $b$ appeared at $F_{\rm act} \simeq 0.1$ that, for the temperature 
 used, $T=0.05$, corresponds to Pe $\simeq 4$.  Thus, $f_A(\mbox{Pe}, \phi)$ should
 be monotonically increasing with Pe, at all fixed $\phi$, 
 as it results  when comparing the data on the different panels in Fig.~\ref{fig:diffusion_cm}.
 In Fig.~\ref{fig:diffusion_cm} we included, with dotted black lines, the exponential fits in Eq.~(\ref{eq:exponential-fit})
 where the only free parameter is $b(\mbox{Pe})$. The values of $b(\mbox{Pe})$ are $
1.1,\ 1.6,\  2.8,\ 4.1$ for 
Pe = $4, \ 20, \ 40, \ 66$, in agreement with what we reported  in~\cite{Suma14b}. 

However, while we see that the exponential fit is very good at all $\phi$ for Pe = $40$ and Pe  = $66$, it is not as good for the smaller
Pe data. The red line-points in Fig.~\ref{fig:diffusion_cm} represent, instead, the result of the fit 
\begin{eqnarray}
D_A(F_{\rm act}, T, \phi) &=& D_A(F_{\rm act}, T, 0) 
\nonumber\\
&& \times  [1+ a_1 (\mbox{Pe}) \ \phi + a_2(\mbox{Pe}) \ \phi^2]
\; . 
\label{eq:quadratic-fit}
\end{eqnarray}
This functional form gives a better representation of the data {than the exponential} for Pe = $4$ and Pe = $20$, 
{which is, in a sense, natural since one expects to recover a rather complex Tokuyama-Oppenheim 
like form in the limit Pe $\to 0$. The exponential and polynomial fits are of
equivalent quality} for Pe = $40$, 
 while {the polynomial fit} is clearly worse than the 
exponential one for Pe = $66$. The fitting parameters are given in the keys. One notices that $a_1$ is negative in all 
cases while $a_2$ changes sign  from negative at Pe $<  20$ to positive at Pe $> 20$ (leading to a growing behaviour 
at large $\phi$ that is not physical). At Pe = 20 the 
density dependence is almost linear as $a_2$ is very close to zero.

\subsection{Rotational diffusion properties}
\label{sec:rotational}

Having discussed in detail the translational diffusion properties we turn now to the 
rotational ones.

{
\subsubsection{Dynamic regimes}
}

\vspace{0.75cm}
\begin{figure*}[ht]
\begin{center}
  \begin{tabular}{cc}
      \includegraphics[scale=0.67]{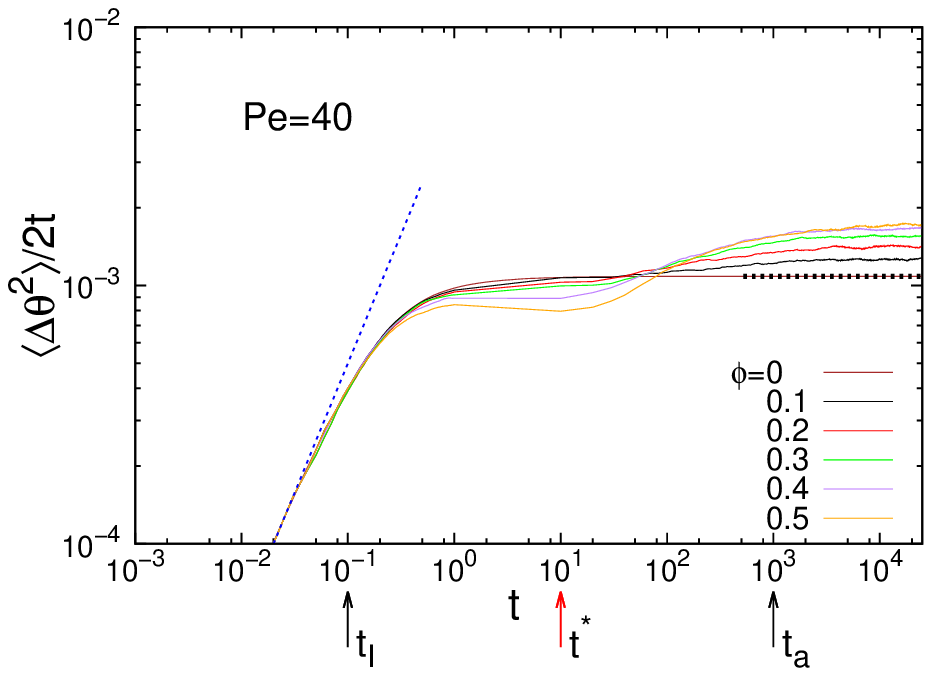}
     \includegraphics[scale=0.67]{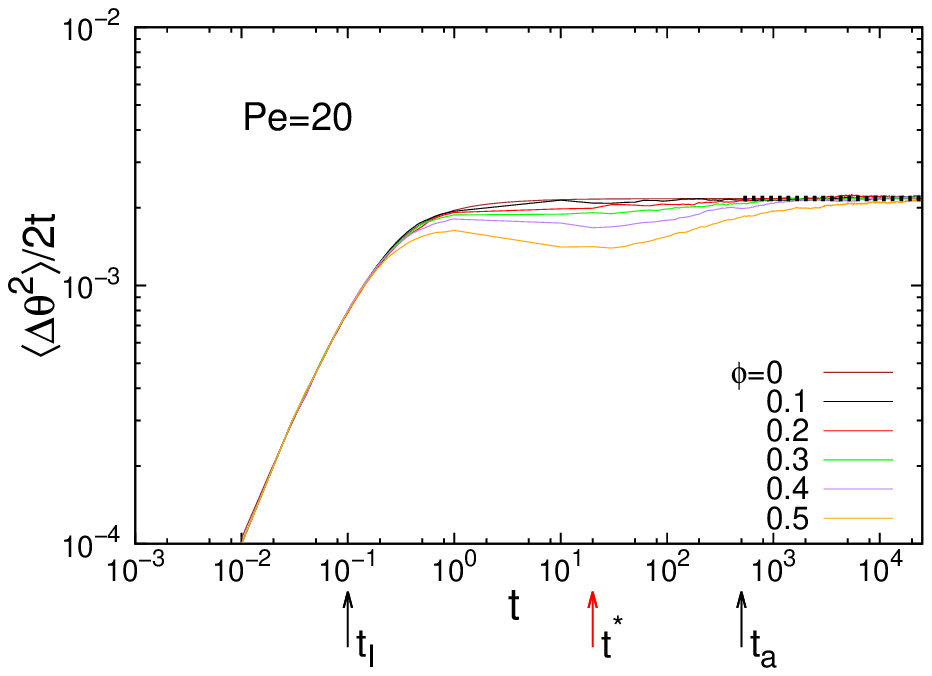}    \\
    \includegraphics[scale=0.67]{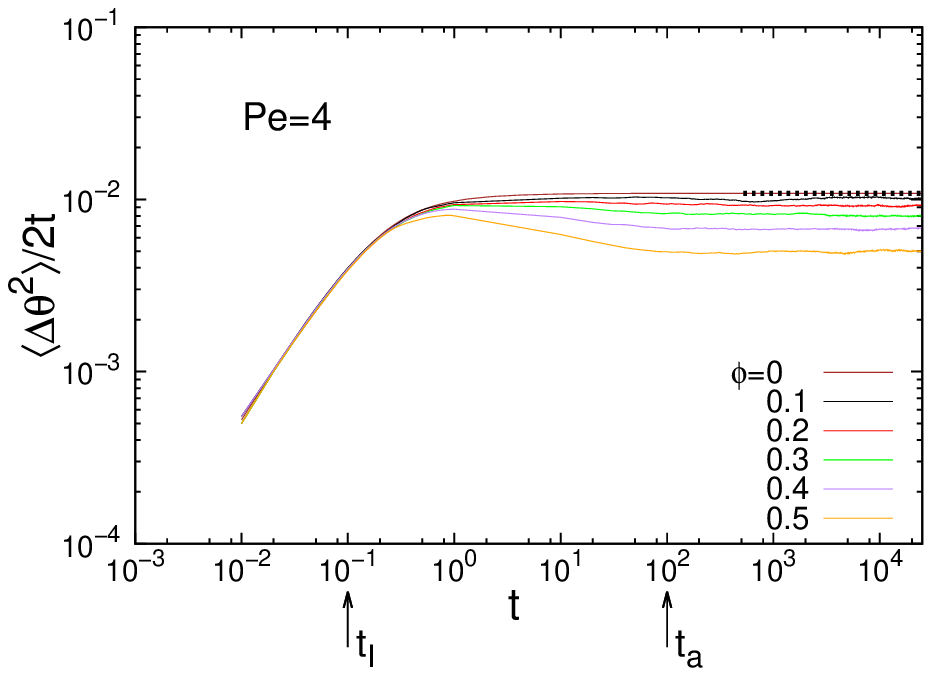}
          \includegraphics[scale=0.67]{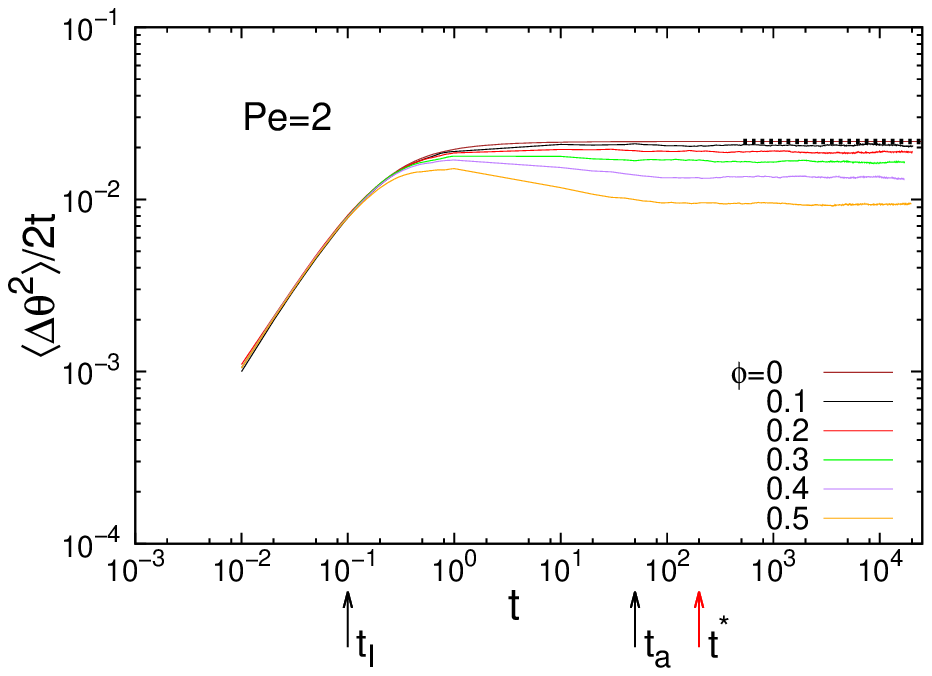}    \\
    \end{tabular}
\caption{The angular MSD for the Pe numbers in the labels and, with different lines, various 
densities given in the keys to each panel.
The dash in the first panel highlights the initial ballistic behavior. The vertical arrows
indicate the characteristic times $t_I, \ t^*,\  t_a$. 
{ 
The horizontal dotted lines at long times correspond to the values 
of the single dumbbell diffusion constant $D_R$ from Eq.~(\ref{eq:DR}).}
 Note that the vertical scale is different in the two
panels above and in the two panels below.
}
\label{fig:mean_square_displacement_angular} 
\end{center}
\end{figure*}

In Fig.~\ref{fig:mean_square_displacement_angular}  we display  the angular MSD 
normalized by time-delay. The four panels 
show data obtained for the same parameters as the ones used in Fig.~\ref{fig:mean_square_displacement_cm}
with  $F_{\rm act}=0.1$.
 Each panel, corresponding to the cases with 
$T=0.005, \ 0.01, \ 0.05, \ 1$  { (Pe = $40, \ 20, \ 4, \ 2$,  respectively),}
includes curves for
five finite densities, $\phi=0.1, \ 0.2, \ 0.3, \ 0.4, \ 0.5$, {and the single dumbbell limit, $\phi=0$, as
labeled in the key}. 
These plots also show several interesting features:\\

\noindent
-- In all cases there is a first ballistic regime with a pre-factor that is independent of $\phi$ and increases with temperature
(The case $t\ll t_I=m_{\rm d}/\gamma$ of the single dumbbell.)
\\
-- Next,  the dynamics slow down and, depending on $T$ and $\phi$, the normalised MSD 
may attain an ever-lasting plateau associated to normal diffusion  for low $\phi$ at any temperature,
or even decrease, suggesting sub-diffusion,
at  high enough $\phi$. 
\\
-- At low temperature $T=0.005, \ 0.01$  (Pe = $40, \ 20$)  
and sufficiently high density the dynamics accelerate next, with a second super-diffusive regime
that crosses over to a final diffusive regime.
\\
-- In the late normal diffusive regime  all curves saturate and the height of the  plateau yields the 
different $D_R$ coefficients that we discuss below.

{
The effect of Pe and $\phi$ are stronger on the rotational MSD  than 
on the translational MSD. New regimes appear in the rotational collective motion
with respect to the individual molecular limit.
In the phase separated regime the dumbbell clusters rotate~\cite{Suma13,Suma14}.
It is possible that strong fluctuations not far from the critical point (Pe = 20, 40)
have an important rotational component than enhances/advects rotational 
diffusion giving rise to an observable contribution to displacement also manifestating itself in the appearing of
new dynamical regimes. }

\vskip 0.5cm
\subsubsection{The late-epochs rotation diffusion coefficient}

\begin{figure*}[ht]
\begin{center}
  \begin{tabular}{cc}
      \includegraphics[scale=0.67]{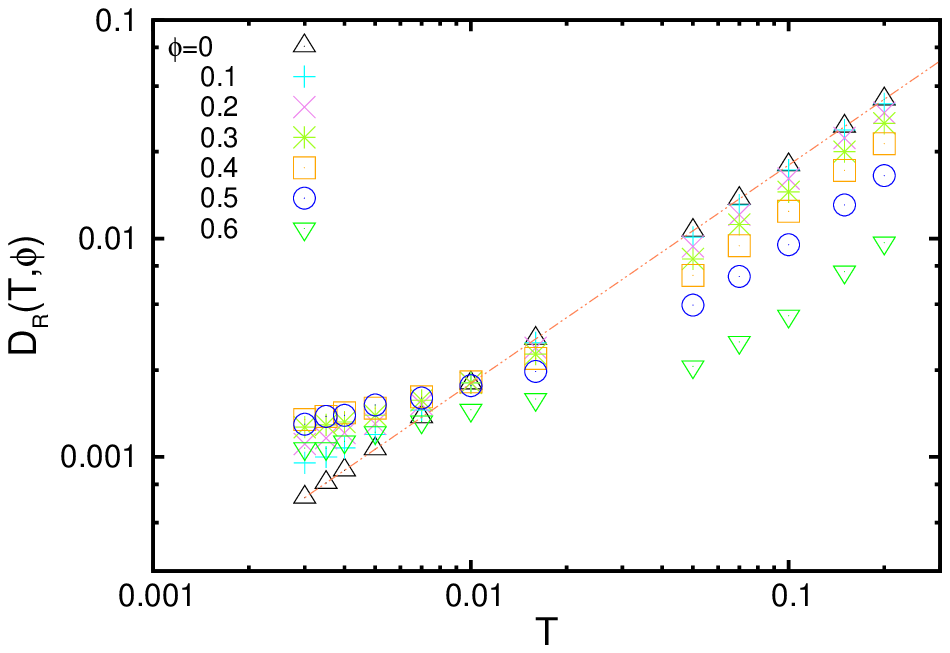}
    \includegraphics[scale=0.67]{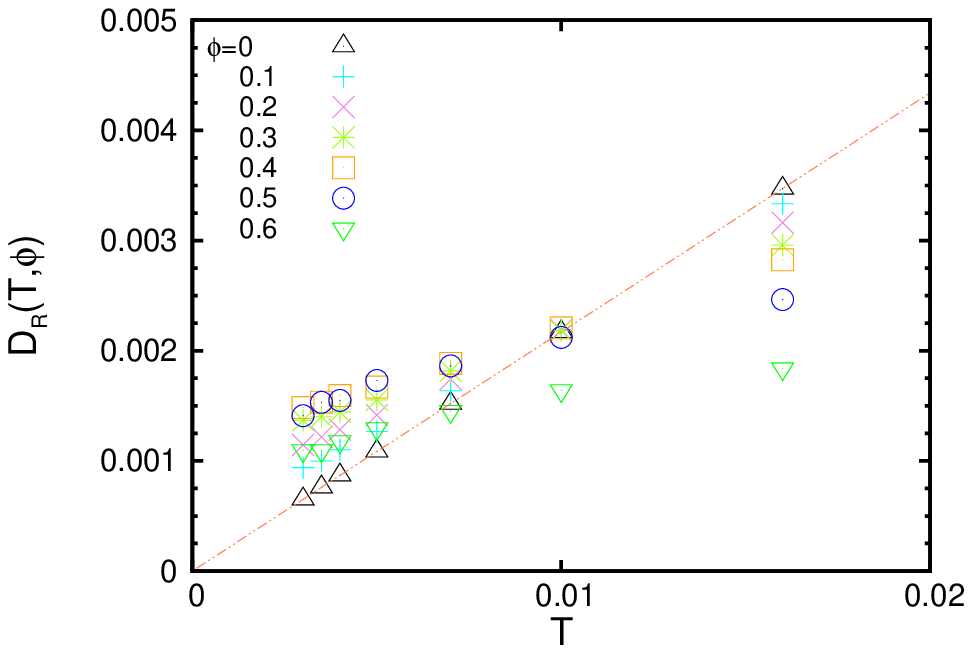}\\ 
    \end{tabular}
\caption{Angular diffusion constant as a function of $T$, for $F_{\rm act}=0.1$ at fixed $\phi$ in logarithmic scale 
(left panel) and a zoom  on the low temperature behaviour in linear scale (right panel). The data points joined with a line 
for $\phi=0$ represent the theoretical expectation (\ref{eq:DR}). 
}
\label{fig:diffusion_angular_Fa0_1_T_dep} 
\end{center}
\end{figure*}

We now study whether the linear temperature dependence of the single dumbbell angular 
diffusion constant, Eq.~(\ref{eq:DR}), survives the interactions between dumbbells in the finite density problem, 
see Fig.~\ref{fig:diffusion_angular_Fa0_1_T_dep}. 
The data points are compatible with a linear behaviour at 
sufficiently high temperature, with a slope that depends upon $\phi$.  The trend in the curves 
reverses below the cross-over 
at $T\simeq 0.01$ with larger values of $D_R$ for larger values of $\phi$ (see the right panel in the same figure). 
 
 \vspace{0.75cm}
 
\begin{figure*}[ht]
\begin{center}
  \begin{tabular}{cc}
      \includegraphics[scale=0.68]{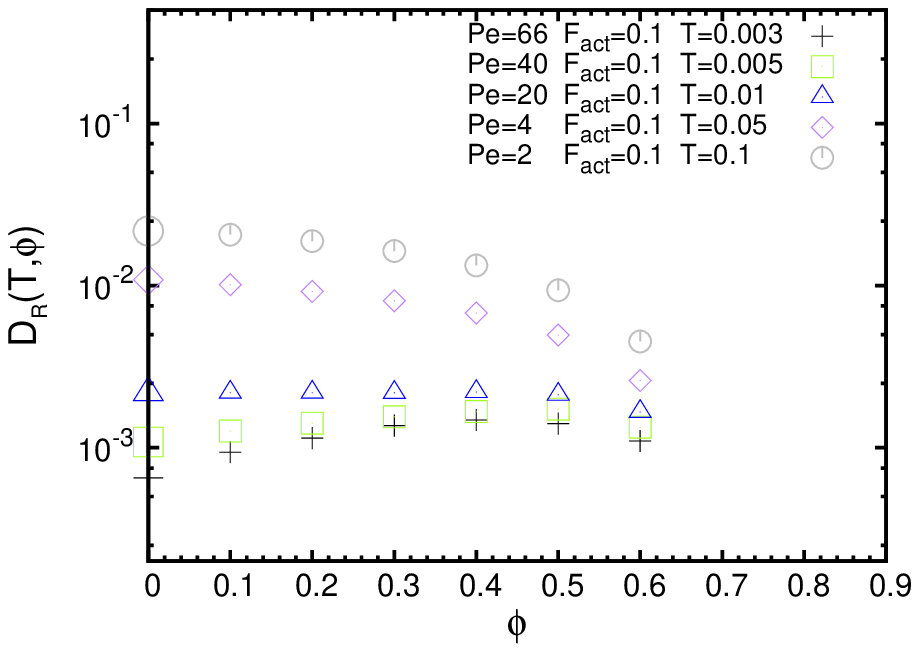}
    \includegraphics[scale=0.68]{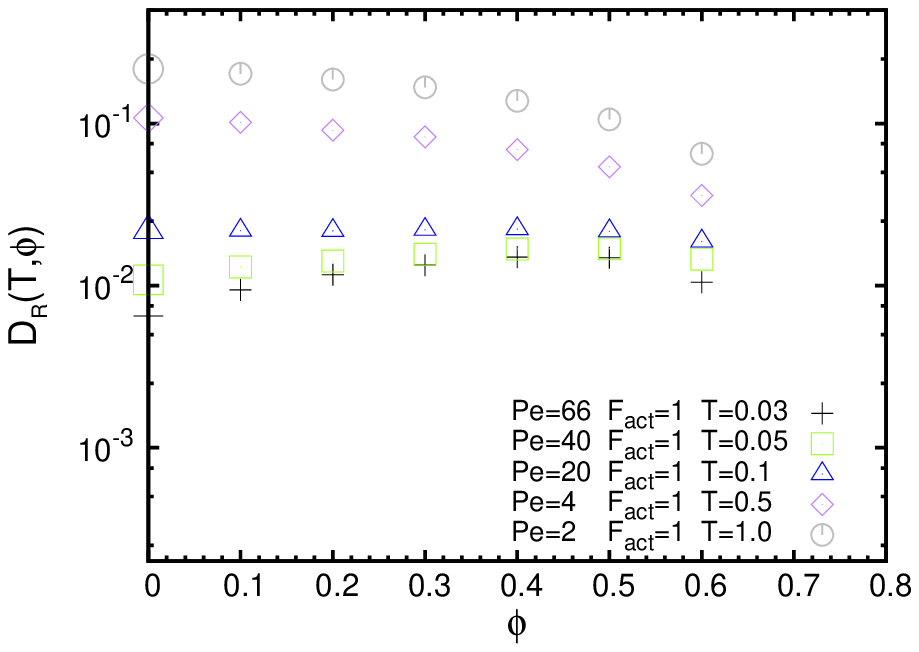}\\  
    \end{tabular}
\caption{Angular diffusion constant for various $T$ under the same active force $F_{\rm act}=0.1$ (left panel) and 
$F_{\rm act} = 1$ (right panel). The only common temperatures on the two panels are $T=0.05,\ 0.1$ and, consistently,
$D_R$ at these temperatures is the same at $\phi=0$. 
}
\label{fig:diffusion_angular_Fa0_1} 
\end{center}
\end{figure*}

From Fig.~\ref{fig:diffusion_angular_Fa0_1} one easily concludes that the $F_{\rm act}$-independence of $D_R$ is lost as soon 
as the interaction between dumbbells is switched on at finite density.
This fact can be seen, for instance, by comparing the $T=0.1$ data, one of the two temperatures included in both panels, 
sharing the same value, slightly larger than $10^{-2}$, at $\phi=0$. 
While in the case $F_{\rm act}=0.1$ (left panel) $D_R$ clearly decreases with $\phi$, in the 
case  $F_{\rm act}=1$ (right panel) $D_R$ is almost constant.  
These figures also show the change in 
trend operated at an $F_{\rm act}$-dependent $T$: at high temperature $D_R$ {\it decreases} with $\phi$ while at 
low temperature $D_R$ {\it increases} with $\phi$. The change occurs at $T\simeq 0.01$ for $F_{\rm act} = 0.1$ and at 
 $T\simeq 0.1$ for $F_{\rm act} = 1$ suggesting that the change is controlled by Pe.

\vspace{0.75cm}

\begin{figure*}[ht]
\begin{center}
  \begin{tabular}{cc}
      \includegraphics[scale=0.67]{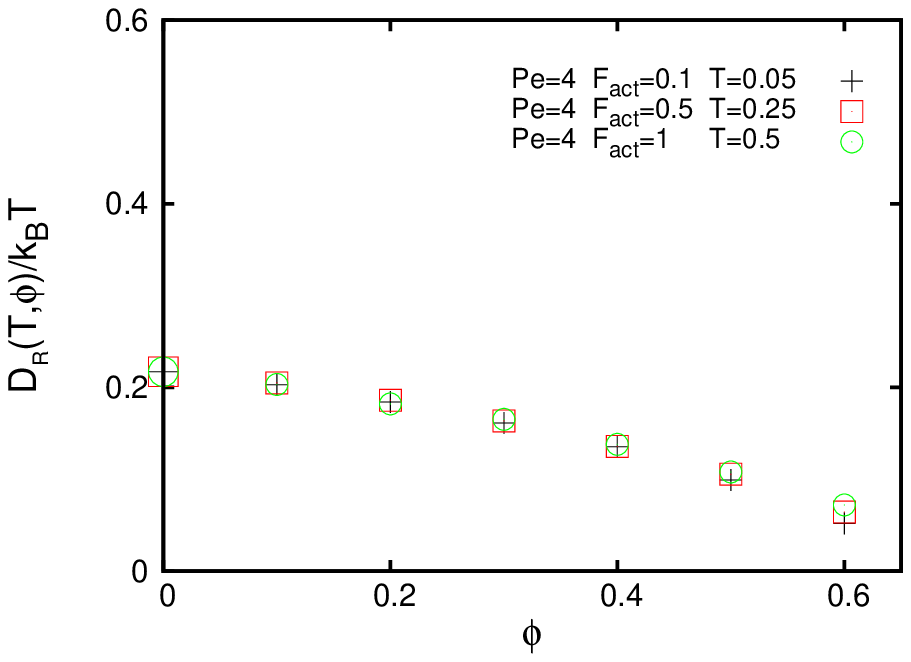}
    \includegraphics[scale=0.67]{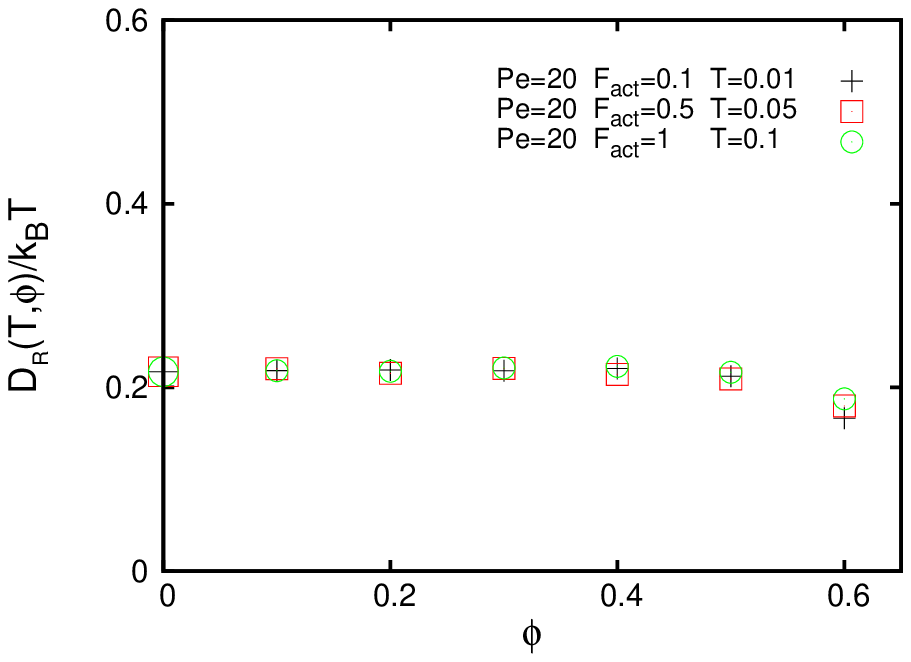}  
    \\
    \includegraphics[scale=0.67]{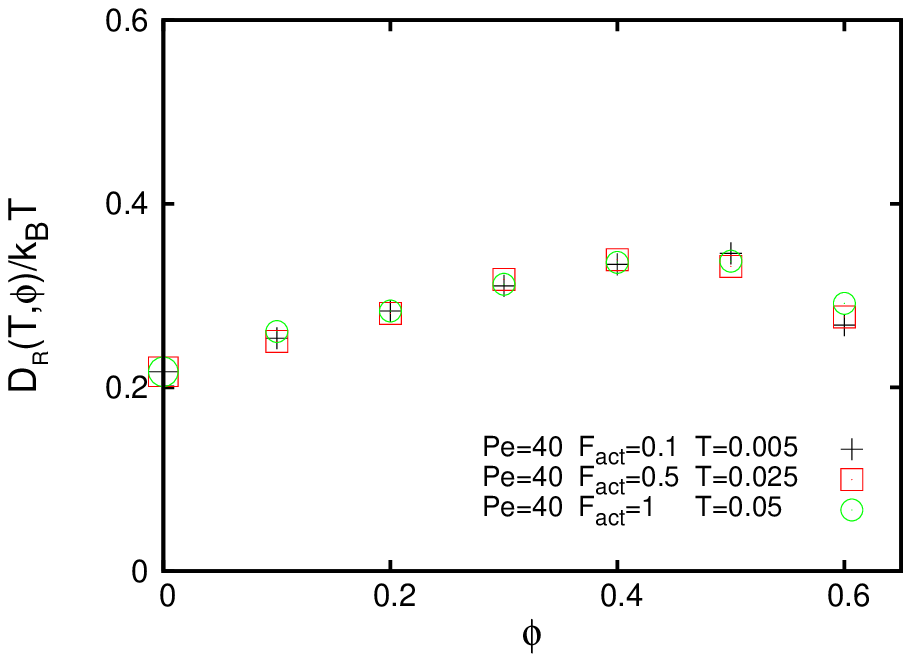}
    \includegraphics[scale=0.67]{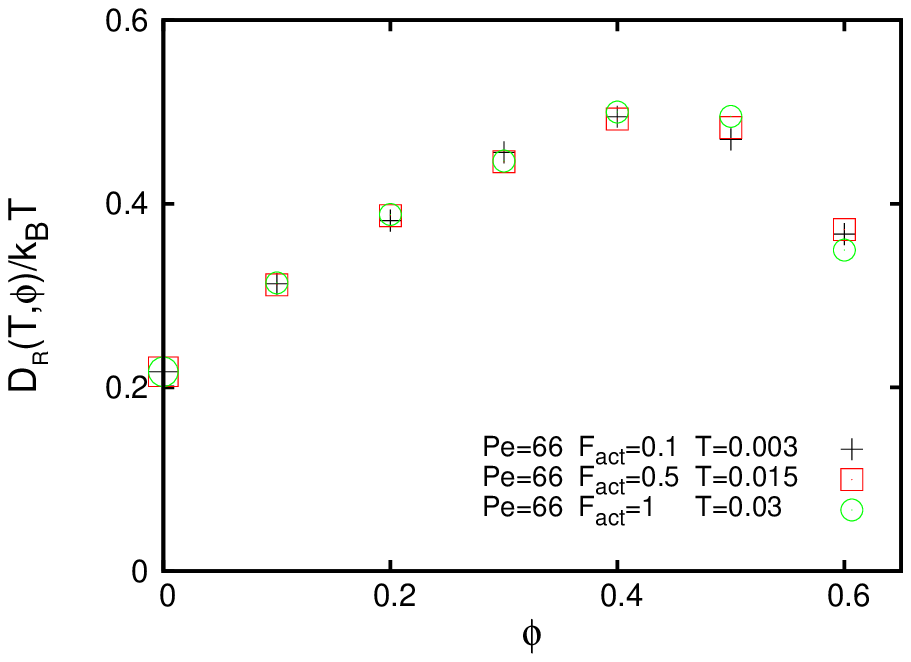}\\
    \end{tabular}
\caption{
Rotational diffusion constant over $k_BT$ for Pe $= 4, \ 20,\ 40, \ 66$ as shown in the keys. For each value of the 
P\'eclet number three couples of values of $F_{\rm act}$ and $T$ were used. 
All master curves take the value $2/(\gamma \sigma_{\rm d}^2) \approx 0.215$ at $\phi=0$.
At small $\phi$, there is a noticeable change in trend at Pe $\simeq 20$. At sufficiently large $\phi$ all curves decrease with increasing 
$\phi$.
}
\label{fig:diffusion_angular_Pe} 
\end{center}
\end{figure*}

Finally, we analyse whether $D_R/(k_BT)$ depends  on $F_{\rm act}$ only {\it via} Pe. To this end, in Fig.~\ref{fig:diffusion_angular_Pe}
we repeat the analysis shown in Fig.~\ref{fig:diffusion_cm} for $D_A$. The four panels show $D_R/(k_BT)$ against
$\phi$ for Pe = $4, \ 20, \ 40, \ 66$. In each panel we include data for three pairs of $F_{\rm act}$ and $T$ leading to the same Pe. 
We see that the data points collapse on different master curves in each panel. 
This suggests
\begin{equation}
D_R(F_{\rm act}, T, \phi) = k_BT \ f_R(\mbox{Pe}, \phi)
\end{equation}
with $f_R(\mbox{Pe},0) = f_R(0,0) = 2/(\gamma \sigma_{\rm d}^2)$. The data also show a change in trend 
of the function $f_R$ at around Pe = 20. At low densities, while the master curve decreases with $\phi$ for Pe $< 20$, 
it becomes flat at Pe $=20$ and it increases with $\phi$ for Pe $> 20$. This would suggest:
\begin{equation}
f_R (\mbox{Pe},\phi) \simeq \frac{2}{\gamma \sigma_{\rm d}^2 } + a(\mbox{Pe},\phi)  \ , 
\end{equation}
with $a($Pe, $\phi)$ almost linear in $\phi$ and the slope changing sign at Pe $\simeq 20$ for small $\phi$. All panels, 
i.e. at all Pe, 
show a cross-over at high enough densities after which the rotational diffusion constant 
decreases with increasing density.

{ A possible explanation of the different density-dependence  of $D_R$ at small and large P\'eclet  can be found from following 
the  evolution of a single tracer dumbbell at intermediate densities, $ \phi \approx 0.4$ for example, 
as it can be seen in the supplementary movies in Ref.~\footnote{See the supplemental movies 1-6 at [URL will be inserted by publisher]. Movies 1-3 refer to the case Pe = 2 ($F_{\rm act}=0.05$, $T=0.05$) and increasing densities $\phi=0.1, 0.4, 0.7$ in order. Movies 4-6 refer to the case of Pe = 40 ($F_{\rm act}=1$, $T=0.05$) and same increasing densities $\phi=0.1, 0.4, 0.7$. A tracer dumbbell is coloured in blue 
to better follow the trajectory of a single particle.}. One observes that at low Pe (Pe = 2) 
the system is very uniformly distributed and the movement of the tracer dumbbell is 
inhibited by  the `cages' formed by  surrounding dumbbells. 
Collisions are frequent but each of them only produces a small
angular displacement. In this case the effect of increasing the density is to decrease
both the rotational and translational diffusion coefficients. 
On the other hand, at high Peclet (Pe = 40), small fluctuating clusters can be observed (their presence is also 
signalled by a peak in the structure factor \cite{Suma14b}). 
This has relevant effects on the behaviour of the 
tracer dumbbell. First, there are  particle depleted regions which are large enough to  allow significant angular displacements without collisions.
Second, angular displacements appear to be  enhanced when the tracer dumbbell meets a cluster and is advected by its motion.  
On the other hand, at still higher densities  the cage effect becomes again preeminent so that 
 rotations are inhibited and $D_R$ decreases.
}{
Note that both $D_R$ and  $D_A$ change behavior at Pe $=20$  (the translational diffusion coefficient 
 $D_A$ is  a convex function of density for Pe $<$ 20 and changes curvature for Pe $>$ 20). 
We find the fact that these cross-overs occur at the same P\'eclet  worth to be stressed
even though it is difficult to argue about its implications.}

\vspace{0.75cm}




\section{Conclusions}
\label{sec:conclusions}

We presented a thorough study of the translational and rotational MSD of a system of 
interacting active dumbbells. We focused on the  regimes where the global 
system is homogeneous. { Higher densities than the ones used in~\cite{wu,Leptos09} have been  considered
with  the P\'eclet number small enough (possibly much smaller than in the experiments) 
to keep the system in the homogenous phase.}

{
We first analysed the single molecule dynamics as a benchmark to later characterize the 
finite density effects. In the passive case, Pe = 0, the 
translational and rotational MSDs
show a standard cross-over from ballistic motion to normal diffusion at the inertial time 
$t_I=m_{\rm d}/\gamma$. { Under the active force, the normal diffusion of the center of mass is accelerated after
a time-scale $t^* \propto t_a/{\mbox{Pe}}^2$  with $t_a=\gamma \sigma_{\rm d}^2/(2k_BT)$
and, still later, after $t_a$, a new diffusive regime is 
reached with a diffusion constant that is enhanced with respect to the one in the passive limit
as a quadratic function of the P\'eclet number. Instead, the rotational properties of the active 
dumbbell are not modified by the longitudinal active force; all torque is exerted by the thermal noise.
}}

{Then we turned to the analysis of the mixed density and active force 
effects on the collective motion of the interacting system.}

{
The rich dynamic structure of the center of mass translational motion 
of the single molecule,}  with the four distinct time regimes {summarized above},
survives under finite densities with modified parameters.
The super-diffusive behaviour shown  in~\cite{wu} {is reminiscent of} the second ballistic 
regime in the interacting active dumbbell system at finite densities.
The diffusion constant  $D_A$ in the last diffusive regime has a non-monotonic dependence on 
temperature, as for the single dumbbell case, and it decreases with increasing self-propelled particle
density at all temperatures. Moreover,  { the ratio $D_A/(k_BT)$ 
depends on temperature and active force only 
through the P\'eclet number at all densities explored.
This ratio, at fixed density, is an increasing function of $\rm Pe$.
 All these results are consistent with those found in our
previous paper~\cite{Suma14b} where it was also shown that the 
ratio between the translational coefficient diffusion at  finite density and the one  for the single dumbbell
had a non-monotonic Pe dependence.
} 

Next we moved to the analysis of the rotational MSD. While in the single dumbbell case its time-delay dependence 
is rather simple, with a single cross-over between ballistic and diffusive behaviour, intermediate regimes, {roughly for $t _I \ll t \ll t^*$ and 
$t^* \ll t \ll t_a$}, appear at 
finite densities. The late epochs diffusion constant  $D_R$ increases with temperature (though not linearly) at all 
densities and active forces simulated. The independence on active force is lost at finite densities. 
The ratio  $D_R/(k_BT)$  
depends on temperature and activity only through the P\'eclet number. 
At low densities, its dependence on density changes from decreasing at low Pe to increasing at high Pe.
{
This  change in behaviour can be related to  the  large scale density fluctuations that appear close to the transition 
from the homogeneous to the aggregated  phase at a critical Pe.
In the aggregated phase  large and rather compact clusters
rotate coherently~\cite{Suma13,Suma14}. Not far from the transition, in the homogenoues phase,
fluctuating clusters with some coherent rotation are observable and these  may be the cause for the 
increase of $D_R$ with $\phi$.} On the other hand, at large enough densities
rotations are strongly inhibited and the value of $D_R$ decreases for all Pe. 

{
The fluctuations of translational and rotational displacements have been characterized in~\cite{Suma15}.
Special emphasis was put on the identification of the regimes in which the fluctuations are non-Gaussian. 
See this reference for more details.
} 

After this work we plan to analyse the motion of tracers in contact with this active sample and, especially,
to analyse the existence of a parameter to be interpreted as an effective temperature from the 
mobility and diffusive properties of the sample and the tracers, in the manner done 
in~\cite{cugl-mossa1,cugl-mossa2,cugl-mossa3,Palacci10,Shen04,Shen05,Wang11,Wang11b,Tailleur09,Szamel14} 
for different active systems.

\vspace{0.5cm}

\noindent \underline{Acknowledgments}:
L. F. C. is a member of Institut Universitaire de France {and 
acknowledges CNRS PICS06691 for financial support}.  G.G. acknowledges the support of  
MIUR (project PRIN 2012NNRKAF).

\vspace{1cm}

\bibliographystyle{apsrev4-1} 
\bibliography{dumbbells-biblio}

\begin{thebibliography}{62}%
\makeatletter
\providecommand \@ifxundefined [1]{%
 \@ifx{#1\undefined}
}%
\providecommand \@ifnum [1]{%
 \ifnum #1\expandafter \@firstoftwo
 \else \expandafter \@secondoftwo
 \fi
}%
\providecommand \@ifx [1]{%
 \ifx #1\expandafter \@firstoftwo
 \else \expandafter \@secondoftwo
 \fi
}%
\providecommand \natexlab [1]{#1}%
\providecommand \enquote  [1]{``#1''}%
\providecommand \bibnamefont  [1]{#1}%
\providecommand \bibfnamefont [1]{#1}%
\providecommand \citenamefont [1]{#1}%
\providecommand \href@noop [0]{\@secondoftwo}%
\providecommand \href [0]{\begingroup \@sanitize@url \@href}%
\providecommand \@href[1]{\@@startlink{#1}\@@href}%
\providecommand \@@href[1]{\endgroup#1\@@endlink}%
\providecommand \@sanitize@url [0]{\catcode `\\12\catcode `\$12\catcode
  `\&12\catcode `\#12\catcode `\^12\catcode `\_12\catcode `\%12\relax}%
\providecommand \@@startlink[1]{}%
\providecommand \@@endlink[0]{}%
\providecommand \url  [0]{\begingroup\@sanitize@url \@url }%
\providecommand \@url [1]{\endgroup\@href {#1}{\urlprefix }}%
\providecommand \urlprefix  [0]{URL }%
\providecommand \Eprint [0]{\href }%
\providecommand \doibase [0]{http://dx.doi.org/}%
\providecommand \selectlanguage [0]{\@gobble}%
\providecommand \bibinfo  [0]{\@secondoftwo}%
\providecommand \bibfield  [0]{\@secondoftwo}%
\providecommand \translation [1]{[#1]}%
\providecommand \BibitemOpen [0]{}%
\providecommand \bibitemStop [0]{}%
\providecommand \bibitemNoStop [0]{.\EOS\space}%
\providecommand \EOS [0]{\spacefactor3000\relax}%
\providecommand \BibitemShut  [1]{\csname bibitem#1\endcsname}%
\let\auto@bib@innerbib\@empty
\bibitem [{\citenamefont {Toner}\ \emph {et~al.}(2005)\citenamefont {Toner},
  \citenamefont {Tu},\ and\ \citenamefont {Ramaswamy}}]{Toner05}%
  \BibitemOpen
  \bibfield  {author} {\bibinfo {author} {\bibfnamefont {J.}~\bibnamefont
  {Toner}}, \bibinfo {author} {\bibfnamefont {Y.}~\bibnamefont {Tu}}, \ and\
  \bibinfo {author} {\bibfnamefont {S.}~\bibnamefont {Ramaswamy}},\ }\href@noop
  {} {\bibfield  {journal} {\bibinfo  {journal} {Ann. of Phys.}\ }\textbf
  {\bibinfo {volume} {318}},\ \bibinfo {pages} {170} (\bibinfo {year}
  {2005})}\BibitemShut {NoStop}%
\bibitem [{\citenamefont {Fletcher}\ and\ \citenamefont
  {Geissler}(2009)}]{Fletcher09}%
  \BibitemOpen
  \bibfield  {author} {\bibinfo {author} {\bibfnamefont {D.~A.}\ \bibnamefont
  {Fletcher}}\ and\ \bibinfo {author} {\bibfnamefont {P.~L.}\ \bibnamefont
  {Geissler}},\ }\href@noop {} {\bibfield  {journal} {\bibinfo  {journal} {Ann.
  Rev. Phys. Chem.}\ }\textbf {\bibinfo {volume} {60}},\ \bibinfo {pages} {469}
  (\bibinfo {year} {2009})}\BibitemShut {NoStop}%
\bibitem [{\citenamefont {Menon}(2010)}]{menon}%
  \BibitemOpen
  \bibfield  {author} {\bibinfo {author} {\bibfnamefont {G.}~\bibnamefont
  {Menon}},\ }in\ \href@noop {} {\emph {\bibinfo {booktitle} {Rheology of
  Complex Fluids}}},\ \bibinfo {editor} {edited by\ \bibinfo {editor}
  {\bibfnamefont {J.}~\bibnamefont {Krishnan}}, \bibinfo {editor}
  {\bibfnamefont {A.}~\bibnamefont {Deshpande}}, \ and\ \bibinfo {editor}
  {\bibfnamefont {P.}~\bibnamefont {Kumar}}}\ (\bibinfo  {publisher}
  {Springer},\ \bibinfo {year} {2010})\BibitemShut {NoStop}%
\bibitem [{\citenamefont {Ramaswamy}(2010)}]{Ramaswamy10}%
  \BibitemOpen
  \bibfield  {author} {\bibinfo {author} {\bibfnamefont {S.}~\bibnamefont
  {Ramaswamy}},\ }\href@noop {} {\bibfield  {journal} {\bibinfo  {journal}
  {Ann. Rev. Cond. Matt. Phys.}\ }\textbf {\bibinfo {volume} {1}},\ \bibinfo
  {pages} {323} (\bibinfo {year} {2010})}\BibitemShut {NoStop}%
\bibitem [{\citenamefont {Cates}(2012)}]{Cates12}%
  \BibitemOpen
  \bibfield  {author} {\bibinfo {author} {\bibfnamefont {M.~E.}\ \bibnamefont
  {Cates}},\ }\href@noop {} {\bibfield  {journal} {\bibinfo  {journal} {Rep.
  Prog. Phys.}\ }\textbf {\bibinfo {volume} {75}},\ \bibinfo {pages} {042601}
  (\bibinfo {year} {2012})}\BibitemShut {NoStop}%
\bibitem [{\citenamefont {Romanczuk}\ \emph {et~al.}(2012)\citenamefont
  {Romanczuk}, \citenamefont {B\"ar}, \citenamefont {Ebeling}, \citenamefont
  {Lindner},\ and\ \citenamefont {Schimansky-Geier}}]{Romanczuk12}%
  \BibitemOpen
  \bibfield  {author} {\bibinfo {author} {\bibfnamefont {P.}~\bibnamefont
  {Romanczuk}}, \bibinfo {author} {\bibfnamefont {M.}~\bibnamefont {B\"ar}},
  \bibinfo {author} {\bibfnamefont {W.}~\bibnamefont {Ebeling}}, \bibinfo
  {author} {\bibfnamefont {B.}~\bibnamefont {Lindner}}, \ and\ \bibinfo
  {author} {\bibfnamefont {L.}~\bibnamefont {Schimansky-Geier}},\ }\href@noop
  {} {\bibfield  {journal} {\bibinfo  {journal} {Eur. Phys. J. Special topics}\
  }\textbf {\bibinfo {volume} {202}},\ \bibinfo {pages} {1} (\bibinfo {year}
  {2012})}\BibitemShut {NoStop}%
\bibitem [{\citenamefont {Vicsek}\ and\ \citenamefont
  {Zafeiris}(2012)}]{Vicsek12}%
  \BibitemOpen
  \bibfield  {author} {\bibinfo {author} {\bibfnamefont {T.}~\bibnamefont
  {Vicsek}}\ and\ \bibinfo {author} {\bibfnamefont {A.}~\bibnamefont
  {Zafeiris}},\ }\href@noop {} {\bibfield  {journal} {\bibinfo  {journal}
  {Phys. Rep.}\ }\textbf {\bibinfo {volume} {517}},\ \bibinfo {pages} {71}
  (\bibinfo {year} {2012})}\BibitemShut {NoStop}%
\bibitem [{\citenamefont {Marchetti}\ \emph {et~al.}(2013)\citenamefont
  {Marchetti}, \citenamefont {Joanny}, \citenamefont {Ramaswamy}, \citenamefont
  {Liverpool}, \citenamefont {Prost}, \citenamefont {Rao},\ and\ \citenamefont
  {Simha}}]{Marchetti13}%
  \BibitemOpen
  \bibfield  {author} {\bibinfo {author} {\bibfnamefont {M.~C.}\ \bibnamefont
  {Marchetti}}, \bibinfo {author} {\bibfnamefont {J.~F.}\ \bibnamefont
  {Joanny}}, \bibinfo {author} {\bibfnamefont {S.}~\bibnamefont {Ramaswamy}},
  \bibinfo {author} {\bibfnamefont {T.~B.}\ \bibnamefont {Liverpool}}, \bibinfo
  {author} {\bibfnamefont {J.}~\bibnamefont {Prost}}, \bibinfo {author}
  {\bibfnamefont {M.}~\bibnamefont {Rao}}, \ and\ \bibinfo {author}
  {\bibfnamefont {R.~A.}\ \bibnamefont {Simha}},\ }\href@noop {} {\bibfield
  {journal} {\bibinfo  {journal} {Rev. Mod. Phys.}\ }\textbf {\bibinfo {volume}
  {85}},\ \bibinfo {pages} {1143} (\bibinfo {year} {2013})}\BibitemShut
  {NoStop}%
\bibitem [{\citenamefont {de~Magistris}\ and\ \citenamefont
  {Marenduzzo}(2015)}]{Marenduzzo14}%
  \BibitemOpen
  \bibfield  {author} {\bibinfo {author} {\bibfnamefont {G.}~\bibnamefont
  {de~Magistris}}\ and\ \bibinfo {author} {\bibfnamefont {D.}~\bibnamefont
  {Marenduzzo}},\ }\href@noop {} {\bibfield  {journal} {\bibinfo  {journal}
  {Physica A}\ }\textbf {\bibinfo {volume} {418}},\ \bibinfo {pages} {65}
  (\bibinfo {year} {2015})}\BibitemShut {NoStop}%
\bibitem [{\citenamefont {Gonnella}\ \emph {et~al.}(2015)\citenamefont
  {Gonnella}, \citenamefont {Marenduzzo}, \citenamefont {Suma},\ and\
  \citenamefont {Tiribocchi}}]{gonnella2015rev}%
  \BibitemOpen
  \bibfield  {author} {\bibinfo {author} {\bibfnamefont {G.}~\bibnamefont
  {Gonnella}}, \bibinfo {author} {\bibfnamefont {D.}~\bibnamefont
  {Marenduzzo}}, \bibinfo {author} {\bibfnamefont {A.}~\bibnamefont {Suma}}, \
  and\ \bibinfo {author} {\bibfnamefont {A.}~\bibnamefont {Tiribocchi}},\
  }\href@noop {} {\bibfield  {journal} {\bibinfo  {journal} {arXiv preprint
  arXiv:1502.02229}\ } (\bibinfo {year} {2015})},\ \bibinfo {note} {to be
  published in "Comptes Rendus de Physique"}\BibitemShut {NoStop}%
\bibitem [{\citenamefont {Walther}\ and\ \citenamefont
  {M{\"u}ller}(2013)}]{Walther}%
  \BibitemOpen
  \bibfield  {author} {\bibinfo {author} {\bibfnamefont {A.}~\bibnamefont
  {Walther}}\ and\ \bibinfo {author} {\bibfnamefont {A.~H.}\ \bibnamefont
  {M{\"u}ller}},\ }\href@noop {} {\bibfield  {journal} {\bibinfo  {journal}
  {Chem. Rev.}\ }\textbf {\bibinfo {volume} {113}},\ \bibinfo {pages} {5194}
  (\bibinfo {year} {2013})}\BibitemShut {NoStop}%
\bibitem [{\citenamefont {Mendelson}\ \emph {et~al.}(1999)\citenamefont
  {Mendelson}, \citenamefont {Bourque}, \citenamefont {Wilkening},
  \citenamefont {Anderson},\ and\ \citenamefont {Watkins}}]{Mendelson99}%
  \BibitemOpen
  \bibfield  {author} {\bibinfo {author} {\bibfnamefont {N.}~\bibnamefont
  {Mendelson}}, \bibinfo {author} {\bibfnamefont {A.}~\bibnamefont {Bourque}},
  \bibinfo {author} {\bibfnamefont {K.}~\bibnamefont {Wilkening}}, \bibinfo
  {author} {\bibfnamefont {K.}~\bibnamefont {Anderson}}, \ and\ \bibinfo
  {author} {\bibfnamefont {J.}~\bibnamefont {Watkins}},\ }\href@noop {}
  {\bibfield  {journal} {\bibinfo  {journal} {J. Bacteriol.}\ }\textbf
  {\bibinfo {volume} {181}},\ \bibinfo {pages} {600} (\bibinfo {year}
  {1999})}\BibitemShut {NoStop}%
\bibitem [{\citenamefont {Wu}\ and\ \citenamefont {Libchaber}(2000)}]{wu}%
  \BibitemOpen
  \bibfield  {author} {\bibinfo {author} {\bibfnamefont {X.-L.}\ \bibnamefont
  {Wu}}\ and\ \bibinfo {author} {\bibfnamefont {A.}~\bibnamefont {Libchaber}},\
  }\href@noop {} {\bibfield  {journal} {\bibinfo  {journal} {Phys. Rev. Lett.}\
  }\textbf {\bibinfo {volume} {84}},\ \bibinfo {pages} {3017} (\bibinfo {year}
  {2000})}\BibitemShut {NoStop}%
\bibitem [{\citenamefont {Dombrowski}\ \emph {et~al.}(2004)\citenamefont
  {Dombrowski}, \citenamefont {Cisneros}, \citenamefont {Chatkaew},
  \citenamefont {Goldstein},\ and\ \citenamefont {Kessler}}]{Dombrowski04}%
  \BibitemOpen
  \bibfield  {author} {\bibinfo {author} {\bibfnamefont {C.}~\bibnamefont
  {Dombrowski}}, \bibinfo {author} {\bibfnamefont {L.}~\bibnamefont
  {Cisneros}}, \bibinfo {author} {\bibfnamefont {S.}~\bibnamefont {Chatkaew}},
  \bibinfo {author} {\bibfnamefont {R.}~\bibnamefont {Goldstein}}, \ and\
  \bibinfo {author} {\bibfnamefont {J.}~\bibnamefont {Kessler}},\ }\href@noop
  {} {\bibfield  {journal} {\bibinfo  {journal} {Phys. Rev. Lett.}\ }\textbf
  {\bibinfo {volume} {93}},\ \bibinfo {pages} {098103} (\bibinfo {year}
  {2004})}\BibitemShut {NoStop}%
\bibitem [{\citenamefont {Hern{\'a}ndez-Ort{\'{\i}}z}\ \emph
  {et~al.}(2005)\citenamefont {Hern{\'a}ndez-Ort{\'{\i}}z}, \citenamefont
  {Stoltz},\ and\ \citenamefont {Graham}}]{Hernandez05}%
  \BibitemOpen
  \bibfield  {author} {\bibinfo {author} {\bibfnamefont {J.~P.}\ \bibnamefont
  {Hern{\'a}ndez-Ort{\'{\i}}z}}, \bibinfo {author} {\bibfnamefont {C.~G.}\
  \bibnamefont {Stoltz}}, \ and\ \bibinfo {author} {\bibfnamefont {M.~D.}\
  \bibnamefont {Graham}},\ }\href@noop {} {\bibfield  {journal} {\bibinfo
  {journal} {Phys. Rev. Lett.}\ }\textbf {\bibinfo {volume} {95}},\ \bibinfo
  {pages} {204501} (\bibinfo {year} {2005})}\BibitemShut {NoStop}%
\bibitem [{\citenamefont {Riedel}\ \emph {et~al.}(2005)\citenamefont {Riedel},
  \citenamefont {Kruse},\ and\ \citenamefont {Howard}}]{Riedel05}%
  \BibitemOpen
  \bibfield  {author} {\bibinfo {author} {\bibfnamefont {I.}~\bibnamefont
  {Riedel}}, \bibinfo {author} {\bibfnamefont {K.}~\bibnamefont {Kruse}}, \
  and\ \bibinfo {author} {\bibfnamefont {J.}~\bibnamefont {Howard}},\
  }\href@noop {} {\bibfield  {journal} {\bibinfo  {journal} {Science}\ }\textbf
  {\bibinfo {volume} {309}},\ \bibinfo {pages} {300} (\bibinfo {year}
  {2005})}\BibitemShut {NoStop}%
\bibitem [{\citenamefont {Sokolov}\ \emph {et~al.}(2007)\citenamefont
  {Sokolov}, \citenamefont {Aranson}, \citenamefont {Kessler},\ and\
  \citenamefont {Goldstein}}]{Sokolov07}%
  \BibitemOpen
  \bibfield  {author} {\bibinfo {author} {\bibfnamefont {A.}~\bibnamefont
  {Sokolov}}, \bibinfo {author} {\bibfnamefont {I.}~\bibnamefont {Aranson}},
  \bibinfo {author} {\bibfnamefont {J.}~\bibnamefont {Kessler}}, \ and\
  \bibinfo {author} {\bibfnamefont {R.}~\bibnamefont {Goldstein}},\ }\href@noop
  {} {\bibfield  {journal} {\bibinfo  {journal} {Phys. Rev. Lett.}\ }\textbf
  {\bibinfo {volume} {98}},\ \bibinfo {pages} {158102} (\bibinfo {year}
  {2007})}\BibitemShut {NoStop}%
\bibitem [{\citenamefont {Zhang}\ \emph {et~al.}(2009)\citenamefont {Zhang},
  \citenamefont {Be'er}, \citenamefont {Smith}, \citenamefont {Florin},\ and\
  \citenamefont {Swinney}}]{Zhang09}%
  \BibitemOpen
  \bibfield  {author} {\bibinfo {author} {\bibfnamefont {H.}~\bibnamefont
  {Zhang}}, \bibinfo {author} {\bibfnamefont {A.}~\bibnamefont {Be'er}},
  \bibinfo {author} {\bibfnamefont {R.}~\bibnamefont {Smith}}, \bibinfo
  {author} {\bibfnamefont {E.-L.}\ \bibnamefont {Florin}}, \ and\ \bibinfo
  {author} {\bibfnamefont {H.}~\bibnamefont {Swinney}},\ }\href@noop {}
  {\bibfield  {journal} {\bibinfo  {journal} {Europhys. Lett.}\ }\textbf
  {\bibinfo {volume} {87}},\ \bibinfo {pages} {48011} (\bibinfo {year}
  {2009})}\BibitemShut {NoStop}%
\bibitem [{\citenamefont {Tailleur}\ and\ \citenamefont
  {Cates}(2008)}]{Tailleur08}%
  \BibitemOpen
  \bibfield  {author} {\bibinfo {author} {\bibfnamefont {J.}~\bibnamefont
  {Tailleur}}\ and\ \bibinfo {author} {\bibfnamefont {M.~E.}\ \bibnamefont
  {Cates}},\ }\href@noop {} {\bibfield  {journal} {\bibinfo  {journal} {Phys.
  Rev. Lett.}\ }\textbf {\bibinfo {volume} {100}},\ \bibinfo {pages} {218103}
  (\bibinfo {year} {2008})}\BibitemShut {NoStop}%
\bibitem [{\citenamefont {Fily}\ and\ \citenamefont
  {Marchetti}(2012)}]{Fily12}%
  \BibitemOpen
  \bibfield  {author} {\bibinfo {author} {\bibfnamefont {Y.}~\bibnamefont
  {Fily}}\ and\ \bibinfo {author} {\bibfnamefont {M.~C.}\ \bibnamefont
  {Marchetti}},\ }\href@noop {} {\bibfield  {journal} {\bibinfo  {journal}
  {Phys. Rev. Lett.}\ }\textbf {\bibinfo {volume} {108}},\ \bibinfo {pages}
  {235702} (\bibinfo {year} {2012})}\BibitemShut {NoStop}%
\bibitem [{\citenamefont {Fily}\ \emph {et~al.}(2014)\citenamefont {Fily},
  \citenamefont {Henkes},\ and\ \citenamefont {Marchetti}}]{Fily14}%
  \BibitemOpen
  \bibfield  {author} {\bibinfo {author} {\bibfnamefont {Y.}~\bibnamefont
  {Fily}}, \bibinfo {author} {\bibfnamefont {S.}~\bibnamefont {Henkes}}, \ and\
  \bibinfo {author} {\bibfnamefont {M.~C.}\ \bibnamefont {Marchetti}},\
  }\href@noop {} {\bibfield  {journal} {\bibinfo  {journal} {Soft Matter}\
  }\textbf {\bibinfo {volume} {10}},\ \bibinfo {pages} {2132} (\bibinfo {year}
  {2014})}\BibitemShut {NoStop}%
\bibitem [{\citenamefont {Redner}\ \emph {et~al.}(2013)\citenamefont {Redner},
  \citenamefont {Hagan},\ and\ \citenamefont {Baskaran}}]{Redner13}%
  \BibitemOpen
  \bibfield  {author} {\bibinfo {author} {\bibfnamefont {G.~S.}\ \bibnamefont
  {Redner}}, \bibinfo {author} {\bibfnamefont {M.~F.}\ \bibnamefont {Hagan}}, \
  and\ \bibinfo {author} {\bibfnamefont {A.}~\bibnamefont {Baskaran}},\
  }\href@noop {} {\bibfield  {journal} {\bibinfo  {journal} {Phys. Rev. Lett.}\
  }\textbf {\bibinfo {volume} {110}},\ \bibinfo {pages} {055701} (\bibinfo
  {year} {2013})}\BibitemShut {NoStop}%
\bibitem [{\citenamefont {Stenhammar}\ \emph {et~al.}(2013)\citenamefont
  {Stenhammar}, \citenamefont {Tiribocchi}, \citenamefont {Allen},
  \citenamefont {Marenduzzo},\ and\ \citenamefont {Cates}}]{Stenhammer13}%
  \BibitemOpen
  \bibfield  {author} {\bibinfo {author} {\bibfnamefont {J.}~\bibnamefont
  {Stenhammar}}, \bibinfo {author} {\bibfnamefont {A.}~\bibnamefont
  {Tiribocchi}}, \bibinfo {author} {\bibfnamefont {R.~J.}\ \bibnamefont
  {Allen}}, \bibinfo {author} {\bibfnamefont {D.}~\bibnamefont {Marenduzzo}}, \
  and\ \bibinfo {author} {\bibfnamefont {M.~E.}\ \bibnamefont {Cates}},\
  }\href@noop {} {\bibfield  {journal} {\bibinfo  {journal} {Phys. Rev. Lett.}\
  }\textbf {\bibinfo {volume} {111}},\ \bibinfo {pages} {145702} (\bibinfo
  {year} {2013})}\BibitemShut {NoStop}%
\bibitem [{\citenamefont {Gonnella}\ \emph {et~al.}(2014)\citenamefont
  {Gonnella}, \citenamefont {Lamura},\ and\ \citenamefont {Suma}}]{Suma13}%
  \BibitemOpen
  \bibfield  {author} {\bibinfo {author} {\bibfnamefont {G.}~\bibnamefont
  {Gonnella}}, \bibinfo {author} {\bibfnamefont {A.}~\bibnamefont {Lamura}}, \
  and\ \bibinfo {author} {\bibfnamefont {A.}~\bibnamefont {Suma}},\ }\href@noop
  {} {\bibfield  {journal} {\bibinfo  {journal} {Int. J. Mod. Phys. C}\
  }\textbf {\bibinfo {volume} {25}},\ \bibinfo {pages} {1441004} (\bibinfo
  {year} {2014})}\BibitemShut {NoStop}%
\bibitem [{\citenamefont {Suma}\ \emph
  {et~al.}(2014{\natexlab{a}})\citenamefont {Suma}, \citenamefont {Marenduzzo},
  \citenamefont {Gonnella},\ and\ \citenamefont {Orlandini}}]{Suma14}%
  \BibitemOpen
  \bibfield  {author} {\bibinfo {author} {\bibfnamefont {A.}~\bibnamefont
  {Suma}}, \bibinfo {author} {\bibfnamefont {D.}~\bibnamefont {Marenduzzo}},
  \bibinfo {author} {\bibfnamefont {G.}~\bibnamefont {Gonnella}}, \ and\
  \bibinfo {author} {\bibfnamefont {E.}~\bibnamefont {Orlandini}},\ }\href@noop
  {} {\bibfield  {journal} {\bibinfo  {journal} {EPL}\ }\textbf {\bibinfo
  {volume} {108}},\ \bibinfo {pages} {56004} (\bibinfo {year}
  {2014}{\natexlab{a}})}\BibitemShut {NoStop}%
\bibitem [{\citenamefont {Levis}\ and\ \citenamefont
  {Berthier}(2014)}]{Levis-Berthier}%
  \BibitemOpen
  \bibfield  {author} {\bibinfo {author} {\bibfnamefont {D.}~\bibnamefont
  {Levis}}\ and\ \bibinfo {author} {\bibfnamefont {L.}~\bibnamefont
  {Berthier}},\ }\href@noop {} {\bibfield  {journal} {\bibinfo  {journal}
  {Phys. Rev. E}\ }\textbf {\bibinfo {volume} {89}},\ \bibinfo {pages} {062301}
  (\bibinfo {year} {2014})}\BibitemShut {NoStop}%
\bibitem [{\citenamefont {Wittkowski}\ \emph {et~al.}(2014)\citenamefont
  {Wittkowski}, \citenamefont {Tiribocchi}, \citenamefont {Stenhammar},
  \citenamefont {Allen}, \citenamefont {Marenduzzo},\ and\ \citenamefont
  {Cates}}]{Stenhammer14}%
  \BibitemOpen
  \bibfield  {author} {\bibinfo {author} {\bibfnamefont {R.}~\bibnamefont
  {Wittkowski}}, \bibinfo {author} {\bibfnamefont {A.}~\bibnamefont
  {Tiribocchi}}, \bibinfo {author} {\bibfnamefont {J.}~\bibnamefont
  {Stenhammar}}, \bibinfo {author} {\bibfnamefont {R.}~\bibnamefont {Allen}},
  \bibinfo {author} {\bibfnamefont {D.}~\bibnamefont {Marenduzzo}}, \ and\
  \bibinfo {author} {\bibfnamefont {M.}~\bibnamefont {Cates}},\ }\href@noop {}
  {\bibfield  {journal} {\bibinfo  {journal} {Nat. Comm.}\ }\textbf {\bibinfo
  {volume} {5}},\ \bibinfo {pages} {4351} (\bibinfo {year} {2014})}\BibitemShut
  {NoStop}%
\bibitem [{\citenamefont {Buttinoni}\ \emph {et~al.}(2013)\citenamefont
  {Buttinoni}, \citenamefont {Bialk\'e}, \citenamefont {K\"ummel},
  \citenamefont {L\"owen}, \citenamefont {Bechinger},\ and\ \citenamefont
  {Speck}}]{Buttinoni13b}%
  \BibitemOpen
  \bibfield  {author} {\bibinfo {author} {\bibfnamefont {I.}~\bibnamefont
  {Buttinoni}}, \bibinfo {author} {\bibfnamefont {J.}~\bibnamefont {Bialk\'e}},
  \bibinfo {author} {\bibfnamefont {F.}~\bibnamefont {K\"ummel}}, \bibinfo
  {author} {\bibfnamefont {H.}~\bibnamefont {L\"owen}}, \bibinfo {author}
  {\bibfnamefont {C.}~\bibnamefont {Bechinger}}, \ and\ \bibinfo {author}
  {\bibfnamefont {T.}~\bibnamefont {Speck}},\ }\href@noop {} {\bibfield
  {journal} {\bibinfo  {journal} {Phys. Rev. Lett.}\ }\textbf {\bibinfo
  {volume} {110}},\ \bibinfo {pages} {238301} (\bibinfo {year}
  {2013})}\BibitemShut {NoStop}%
\bibitem [{\citenamefont {Palacci}\ \emph {et~al.}(2010)\citenamefont
  {Palacci}, \citenamefont {Cottin-Bizonne}, \citenamefont {Ybert},\ and\
  \citenamefont {Bocquet}}]{Palacci10}%
  \BibitemOpen
  \bibfield  {author} {\bibinfo {author} {\bibfnamefont {J.}~\bibnamefont
  {Palacci}}, \bibinfo {author} {\bibfnamefont {C.}~\bibnamefont
  {Cottin-Bizonne}}, \bibinfo {author} {\bibfnamefont {C.}~\bibnamefont
  {Ybert}}, \ and\ \bibinfo {author} {\bibfnamefont {L.}~\bibnamefont
  {Bocquet}},\ }\href@noop {} {\bibfield  {journal} {\bibinfo  {journal} {Phys.
  Rev. Lett.}\ }\textbf {\bibinfo {volume} {105}},\ \bibinfo {pages} {088304}
  (\bibinfo {year} {2010})}\BibitemShut {NoStop}%
\bibitem [{\citenamefont {Leptos}\ \emph {et~al.}(2009)\citenamefont {Leptos},
  \citenamefont {Guasto}, \citenamefont {Gollub}, \citenamefont {Pesci},\ and\
  \citenamefont {Goldstein}}]{Leptos09}%
  \BibitemOpen
  \bibfield  {author} {\bibinfo {author} {\bibfnamefont {K.~C.}\ \bibnamefont
  {Leptos}}, \bibinfo {author} {\bibfnamefont {J.}~\bibnamefont {Guasto}},
  \bibinfo {author} {\bibfnamefont {J.}~\bibnamefont {Gollub}}, \bibinfo
  {author} {\bibfnamefont {A.~I.}\ \bibnamefont {Pesci}}, \ and\ \bibinfo
  {author} {\bibfnamefont {R.}~\bibnamefont {Goldstein}},\ }\href@noop {}
  {\bibfield  {journal} {\bibinfo  {journal} {Phys. Rev. Lett.}\ }\textbf
  {\bibinfo {volume} {103}},\ \bibinfo {pages} {198103} (\bibinfo {year}
  {2009})}\BibitemShut {NoStop}%
\bibitem [{\citenamefont {Kurtuldu}\ \emph {et~al.}(2011)\citenamefont
  {Kurtuldu}, \citenamefont {Guasto}, \citenamefont {Johnson},\ and\
  \citenamefont {Gollub}}]{Kurtuldu11}%
  \BibitemOpen
  \bibfield  {author} {\bibinfo {author} {\bibfnamefont {H.}~\bibnamefont
  {Kurtuldu}}, \bibinfo {author} {\bibfnamefont {J.}~\bibnamefont {Guasto}},
  \bibinfo {author} {\bibfnamefont {K.}~\bibnamefont {Johnson}}, \ and\
  \bibinfo {author} {\bibfnamefont {J.}~\bibnamefont {Gollub}},\ }\href@noop {}
  {\bibfield  {journal} {\bibinfo  {journal} {Proc. Nat. Acad. Sc.}\ }\textbf
  {\bibinfo {volume} {108}},\ \bibinfo {pages} {10391} (\bibinfo {year}
  {2011})}\BibitemShut {NoStop}%
\bibitem [{\citenamefont {Kasyap}\ \emph {et~al.}(2014)\citenamefont {Kasyap},
  \citenamefont {Koch},\ and\ \citenamefont {Wu}}]{Kasyap14}%
  \BibitemOpen
  \bibfield  {author} {\bibinfo {author} {\bibfnamefont {T.}~\bibnamefont
  {Kasyap}}, \bibinfo {author} {\bibfnamefont {D.}~\bibnamefont {Koch}}, \ and\
  \bibinfo {author} {\bibfnamefont {M.}~\bibnamefont {Wu}},\ }\href@noop {}
  {\bibfield  {journal} {\bibinfo  {journal} {Phys. of Fluids}\ }\textbf
  {\bibinfo {volume} {26}},\ \bibinfo {pages} {081901} (\bibinfo {year}
  {2014})}\BibitemShut {NoStop}%
\bibitem [{\citenamefont {Pushkin}\ and\ \citenamefont
  {Yeomans}(2014)}]{Pushkin14}%
  \BibitemOpen
  \bibfield  {author} {\bibinfo {author} {\bibfnamefont {D.}~\bibnamefont
  {Pushkin}}\ and\ \bibinfo {author} {\bibfnamefont {J.}~\bibnamefont
  {Yeomans}},\ }\href@noop {} {\bibfield  {journal} {\bibinfo  {journal} {J.
  Stat. Mech.}\ ,\ \bibinfo {pages} {P04030}} (\bibinfo {year}
  {2014})}\BibitemShut {NoStop}%
\bibitem [{\citenamefont {Morozov}\ and\ \citenamefont
  {Marenduzzo}(2014)}]{Morozov14}%
  \BibitemOpen
  \bibfield  {author} {\bibinfo {author} {\bibfnamefont {A.}~\bibnamefont
  {Morozov}}\ and\ \bibinfo {author} {\bibfnamefont {D.}~\bibnamefont
  {Marenduzzo}},\ }\href@noop {} {\bibfield  {journal} {\bibinfo  {journal}
  {Soft Matter}\ }\textbf {\bibinfo {volume} {10}},\ \bibinfo {pages} {2748}
  (\bibinfo {year} {2014})}\BibitemShut {NoStop}%
\bibitem [{\citenamefont {Mi{\~n}o}\ \emph {et~al.}(2011)\citenamefont
  {Mi{\~n}o}, \citenamefont {Mallouk}, \citenamefont {Darnige}, \citenamefont
  {Hoyos}, \citenamefont {Dauchet}, \citenamefont {Dunstan}, \citenamefont
  {Soto}, \citenamefont {Wang}, \citenamefont {Rousselet},\ and\ \citenamefont
  {Clement}}]{Mino11}%
  \BibitemOpen
  \bibfield  {author} {\bibinfo {author} {\bibfnamefont {G.}~\bibnamefont
  {Mi{\~n}o}}, \bibinfo {author} {\bibfnamefont {T.~E.}\ \bibnamefont
  {Mallouk}}, \bibinfo {author} {\bibfnamefont {T.}~\bibnamefont {Darnige}},
  \bibinfo {author} {\bibfnamefont {M.}~\bibnamefont {Hoyos}}, \bibinfo
  {author} {\bibfnamefont {J.}~\bibnamefont {Dauchet}}, \bibinfo {author}
  {\bibfnamefont {J.}~\bibnamefont {Dunstan}}, \bibinfo {author} {\bibfnamefont
  {R.}~\bibnamefont {Soto}}, \bibinfo {author} {\bibfnamefont {Y.}~\bibnamefont
  {Wang}}, \bibinfo {author} {\bibfnamefont {A.}~\bibnamefont {Rousselet}}, \
  and\ \bibinfo {author} {\bibfnamefont {E.}~\bibnamefont {Clement}},\
  }\href@noop {} {\bibfield  {journal} {\bibinfo  {journal} {Phys. Rev. Lett.}\
  }\textbf {\bibinfo {volume} {106}},\ \bibinfo {pages} {048102} (\bibinfo
  {year} {2011})}\BibitemShut {NoStop}%
\bibitem [{\citenamefont {Llopis}\ and\ \citenamefont
  {Pagonabarraga}(2006)}]{Llopis06}%
  \BibitemOpen
  \bibfield  {author} {\bibinfo {author} {\bibfnamefont {I.}~\bibnamefont
  {Llopis}}\ and\ \bibinfo {author} {\bibfnamefont {I.}~\bibnamefont
  {Pagonabarraga}},\ }\href@noop {} {\bibfield  {journal} {\bibinfo  {journal}
  {EPL}\ }\textbf {\bibinfo {volume} {999}},\ \bibinfo {pages} {75} (\bibinfo
  {year} {2006})}\BibitemShut {NoStop}%
\bibitem [{\citenamefont {Gr\'egoire}\ and\ \citenamefont
  {Chat\'e}(2001)}]{Gregoire01}%
  \BibitemOpen
  \bibfield  {author} {\bibinfo {author} {\bibfnamefont {G.}~\bibnamefont
  {Gr\'egoire}}\ and\ \bibinfo {author} {\bibfnamefont {Y.}~\bibnamefont
  {Chat\'e}, \bibfnamefont {H.~Tu}},\ }\href@noop {} {\bibfield  {journal}
  {\bibinfo  {journal} {Phys. Rev. E}\ }\textbf {\bibinfo {volume} {64}},\
  \bibinfo {pages} {011902} (\bibinfo {year} {2001})}\BibitemShut {NoStop}%
\bibitem [{\citenamefont {Valeriani}\ \emph {et~al.}(2011)\citenamefont
  {Valeriani}, \citenamefont {Li}, \citenamefont {Novosel}, \citenamefont
  {Arlt},\ and\ \citenamefont {Marenduzzo}}]{valeriani2011colloids}%
  \BibitemOpen
  \bibfield  {author} {\bibinfo {author} {\bibfnamefont {C.}~\bibnamefont
  {Valeriani}}, \bibinfo {author} {\bibfnamefont {M.}~\bibnamefont {Li}},
  \bibinfo {author} {\bibfnamefont {J.}~\bibnamefont {Novosel}}, \bibinfo
  {author} {\bibfnamefont {J.}~\bibnamefont {Arlt}}, \ and\ \bibinfo {author}
  {\bibfnamefont {D.}~\bibnamefont {Marenduzzo}},\ }\href@noop {} {\bibfield
  {journal} {\bibinfo  {journal} {Soft Matter}\ }\textbf {\bibinfo {volume}
  {7}},\ \bibinfo {pages} {5228} (\bibinfo {year} {2011})}\BibitemShut
  {NoStop}%
\bibitem [{\citenamefont {Suma}\ \emph
  {et~al.}(2014{\natexlab{b}})\citenamefont {Suma}, \citenamefont {Gonnella},
  \citenamefont {Laghezza}, \citenamefont {Lamura}, \citenamefont {Mossa},\
  and\ \citenamefont {Cugliandolo}}]{Suma14b}%
  \BibitemOpen
  \bibfield  {author} {\bibinfo {author} {\bibfnamefont {A.}~\bibnamefont
  {Suma}}, \bibinfo {author} {\bibfnamefont {G.}~\bibnamefont {Gonnella}},
  \bibinfo {author} {\bibfnamefont {G.}~\bibnamefont {Laghezza}}, \bibinfo
  {author} {\bibfnamefont {A.}~\bibnamefont {Lamura}}, \bibinfo {author}
  {\bibfnamefont {A.}~\bibnamefont {Mossa}}, \ and\ \bibinfo {author}
  {\bibfnamefont {L.~F.}\ \bibnamefont {Cugliandolo}},\ }\href@noop {}
  {\bibfield  {journal} {\bibinfo  {journal} {Phys. Rev. E}\ }\textbf {\bibinfo
  {volume} {90}},\ \bibinfo {pages} {052130} (\bibinfo {year}
  {2014}{\natexlab{b}})}\BibitemShut {NoStop}%
\bibitem [{\citenamefont {Cugliandolo}(2011)}]{cugl:review}%
  \BibitemOpen
  \bibfield  {author} {\bibinfo {author} {\bibfnamefont {L.~F.}\ \bibnamefont
  {Cugliandolo}},\ }\href@noop {} {\bibfield  {journal} {\bibinfo  {journal}
  {J. Phys. A: Math. and Theor.}\ }\textbf {\bibinfo {volume} {44}},\ \bibinfo
  {pages} {483001} (\bibinfo {year} {2011})}\BibitemShut {NoStop}%
\bibitem [{\citenamefont {Weeks}\ \emph {et~al.}(1971)\citenamefont {Weeks},
  \citenamefont {Chandler},\ and\ \citenamefont {Andersen}}]{Weeks}%
  \BibitemOpen
  \bibfield  {author} {\bibinfo {author} {\bibfnamefont {J.~D.}\ \bibnamefont
  {Weeks}}, \bibinfo {author} {\bibfnamefont {D.}~\bibnamefont {Chandler}}, \
  and\ \bibinfo {author} {\bibfnamefont {H.~C.}\ \bibnamefont {Andersen}},\
  }\href@noop {} {\bibfield  {journal} {\bibinfo  {journal} {J. Chem. Phys.}\
  }\textbf {\bibinfo {volume} {54}},\ \bibinfo {pages} {5237} (\bibinfo {year}
  {1971})}\BibitemShut {NoStop}%
\bibitem [{Note1()}]{Note1}%
  \BibitemOpen
  \bibinfo {note} {In a system with momentum conservation the total force on a
  neutrally buoyant swimmer should indeed be zero. However Brownian dynamics
  theories and simulations neglect fluid-mediated interactions so the only way
  to propel a particle is to apply a force along its direction.}\BibitemShut
  {Stop}%
\bibitem [{\citenamefont {Baskaran}\ and\ \citenamefont
  {Marchetti}(2010)}]{Baskaran10}%
  \BibitemOpen
  \bibfield  {author} {\bibinfo {author} {\bibfnamefont {A.}~\bibnamefont
  {Baskaran}}\ and\ \bibinfo {author} {\bibfnamefont {M.~C.}\ \bibnamefont
  {Marchetti}},\ }\href@noop {} {\bibfield  {journal} {\bibinfo  {journal} {J.
  Stat. Mec.}\ ,\ \bibinfo {pages} {P04019}} (\bibinfo {year}
  {2010})}\BibitemShut {NoStop}%
\bibitem [{\citenamefont {{\O}ksendhal}(2000)}]{Oksendhal}%
  \BibitemOpen
  \bibfield  {author} {\bibinfo {author} {\bibfnamefont {B.}~\bibnamefont
  {{\O}ksendhal}},\ }\href@noop {} {\emph {\bibinfo {title} {Stochastic
  differential equations}}}\ (\bibinfo  {publisher} {Springer-Verlag},\
  \bibinfo {address} {Berlin},\ \bibinfo {year} {2000})\BibitemShut {NoStop}%
\bibitem [{Note2()}]{Note2}%
  \BibitemOpen
  \bibinfo {note} {The use of Stratonovich calculation is quite natural in this
  context, as stressed by van Kampen and others \cite {van1981ito}, as for most
  of physical problems. We can also observe that, following the Ito
  approach~\cite {Gardiner}, the equations of motion for the polar coordinates
  can be written, neglecting the inertial contribution, as $\gamma \protect
  \mathaccentV {dot}05Fr = 2 F_{\protect \rm int} + \protect \frac
  {2k_BT}{r}+\zeta _r$, $\protect \mathaccentV {dot}05F\theta = \protect \frac
  {1}{\gamma r} \zeta _\theta $ with $\zeta _r$, $\zeta _\theta $ Gaussian
  white noises satisfying the same properties as $\zeta _x$, $\zeta _y$ of
  Eq.~(\ref {eq:corr_rum}). This set of equations give the same dynamical
  equations for the momenta Eqs. (\ref {eq:rrr}-\ref {eq:thetptoquadr})
  resulting from Stratonovich approach.}\BibitemShut {Stop}%
\bibitem [{\citenamefont {Gardiner}(1996)}]{Gardiner}%
  \BibitemOpen
  \bibfield  {author} {\bibinfo {author} {\bibfnamefont {C.~W.}\ \bibnamefont
  {Gardiner}},\ }\href@noop {} {\emph {\bibinfo {title} {Handbook of stochastic
  methods for physics, chemistry and the natural sciences}}}\ (\bibinfo
  {publisher} {Springer-Verlag},\ \bibinfo {address} {Berlin Heidelberg},\
  \bibinfo {year} {1996})\BibitemShut {NoStop}%
\bibitem [{\citenamefont {Coffey}\ \emph {et~al.}(2012)\citenamefont {Coffey},
  \citenamefont {Kalmykov},\ and\ \citenamefont {Waldron}}]{Coffey}%
  \BibitemOpen
  \bibfield  {author} {\bibinfo {author} {\bibfnamefont {W.~T.}\ \bibnamefont
  {Coffey}}, \bibinfo {author} {\bibfnamefont {Y.~P.}\ \bibnamefont
  {Kalmykov}}, \ and\ \bibinfo {author} {\bibfnamefont {J.~T.}\ \bibnamefont
  {Waldron}},\ }\href@noop {} {\emph {\bibinfo {title} {The Langevin equation -
  3rd edition}}},\ \bibinfo {series} {World Scientific series in contemporary
  chemical physics}, Vol.~\bibinfo {volume} {27}\ (\bibinfo  {publisher} {World
  Scientific},\ \bibinfo {address} {Singapore},\ \bibinfo {year}
  {2012})\BibitemShut {NoStop}%
\bibitem [{Note3()}]{Note3}%
  \BibitemOpen
  \bibinfo {note} {{Equation~(\ref {eq:rrr}) with the l.h.s. set to zero and
  the potential parameters that we use in the simulations yields $r\approx 0.96
  \sigma _{\protect \rm d}$ quite independently of temperature in the range
  $k_BT \in [10^{-5}, 1]$. In the simulations we find that the fluctuations
  around this value increase weakly with increasing temperature.}}\BibitemShut
  {Stop}%
\bibitem [{\citenamefont {ten Hagen}\ \emph {et~al.}(2011)\citenamefont {ten
  Hagen}, \citenamefont {van Teeffelen},\ and\ \citenamefont
  {L\"owen}}]{Lowen11}%
  \BibitemOpen
  \bibfield  {author} {\bibinfo {author} {\bibfnamefont {B.}~\bibnamefont {ten
  Hagen}}, \bibinfo {author} {\bibfnamefont {S.}~\bibnamefont {van Teeffelen}},
  \ and\ \bibinfo {author} {\bibfnamefont {H.}~\bibnamefont {L\"owen}},\
  }\href@noop {} {\bibfield  {journal} {\bibinfo  {journal} {J. Phys.: Condens.
  Matter}\ }\textbf {\bibinfo {volume} {23}},\ \bibinfo {pages} {194119}
  (\bibinfo {year} {2011})}\BibitemShut {NoStop}%
\bibitem [{\citenamefont {Tokuyama}\ and\ \citenamefont
  {Oppenheim}(1994)}]{Tokuyama}%
  \BibitemOpen
  \bibfield  {author} {\bibinfo {author} {\bibfnamefont {M.}~\bibnamefont
  {Tokuyama}}\ and\ \bibinfo {author} {\bibfnamefont {I.}~\bibnamefont
  {Oppenheim}},\ }\href@noop {} {\bibfield  {journal} {\bibinfo  {journal}
  {Phys. Rev. E}\ }\textbf {\bibinfo {volume} {50}},\ \bibinfo {pages} {16}
  (\bibinfo {year} {1994})}\BibitemShut {NoStop}%
\bibitem [{Note4()}]{Note4}%
  \BibitemOpen
  \bibinfo {note} {See the supplemental movies 1-6 at [URL will be inserted by
  publisher]. Movies 1-3 refer to the case Pe = 2 ($F_{\protect \rm act}=0.05$,
  $T=0.05$) and increasing densities $\phi =0.1, 0.4, 0.7$ in order. Movies 4-6
  refer to the case of Pe = 40 ($F_{\protect \rm act}=1$, $T=0.05$) and same
  increasing densities $\phi =0.1, 0.4, 0.7$. A tracer dumbbell is coloured in
  blue to better follow the trajectory of a single particle.}\BibitemShut
  {Stop}%
\bibitem [{\citenamefont {Cugliandolo}\ \emph {et~al.}(2015)\citenamefont
  {Cugliandolo}, \citenamefont {Gonnella},\ and\ \citenamefont
  {Suma}}]{Suma15}%
  \BibitemOpen
  \bibfield  {author} {\bibinfo {author} {\bibfnamefont {L.~F.}\ \bibnamefont
  {Cugliandolo}}, \bibinfo {author} {\bibfnamefont {G.}~\bibnamefont
  {Gonnella}}, \ and\ \bibinfo {author} {\bibfnamefont {A.}~\bibnamefont
  {Suma}},\ }\href@noop {} {\bibfield  {journal} {\bibinfo  {journal} {Chaos
  and solitons - to appear, arXiv:1504.03549}\ } (\bibinfo {year}
  {2015})}\BibitemShut {NoStop}%
\bibitem [{\citenamefont {Loi}\ \emph {et~al.}(2008)\citenamefont {Loi},
  \citenamefont {Mossa},\ and\ \citenamefont {Cugliandolo}}]{cugl-mossa1}%
  \BibitemOpen
  \bibfield  {author} {\bibinfo {author} {\bibfnamefont {D.}~\bibnamefont
  {Loi}}, \bibinfo {author} {\bibfnamefont {S.}~\bibnamefont {Mossa}}, \ and\
  \bibinfo {author} {\bibfnamefont {L.~F.}\ \bibnamefont {Cugliandolo}},\
  }\href@noop {} {\bibfield  {journal} {\bibinfo  {journal} {Phys. Rev. E}\
  }\textbf {\bibinfo {volume} {77}},\ \bibinfo {pages} {051111} (\bibinfo
  {year} {2008})}\BibitemShut {NoStop}%
\bibitem [{\citenamefont {Loi}\ \emph {et~al.}(2011{\natexlab{a}})\citenamefont
  {Loi}, \citenamefont {Mossa},\ and\ \citenamefont
  {Cugliandolo}}]{cugl-mossa2}%
  \BibitemOpen
  \bibfield  {author} {\bibinfo {author} {\bibfnamefont {D.}~\bibnamefont
  {Loi}}, \bibinfo {author} {\bibfnamefont {S.}~\bibnamefont {Mossa}}, \ and\
  \bibinfo {author} {\bibfnamefont {L.~F.}\ \bibnamefont {Cugliandolo}},\
  }\href@noop {} {\bibfield  {journal} {\bibinfo  {journal} {Soft Matter}\
  }\textbf {\bibinfo {volume} {7}},\ \bibinfo {pages} {3726} (\bibinfo {year}
  {2011}{\natexlab{a}})}\BibitemShut {NoStop}%
\bibitem [{\citenamefont {Loi}\ \emph {et~al.}(2011{\natexlab{b}})\citenamefont
  {Loi}, \citenamefont {Mossa},\ and\ \citenamefont
  {Cugliandolo}}]{cugl-mossa3}%
  \BibitemOpen
  \bibfield  {author} {\bibinfo {author} {\bibfnamefont {D.}~\bibnamefont
  {Loi}}, \bibinfo {author} {\bibfnamefont {S.}~\bibnamefont {Mossa}}, \ and\
  \bibinfo {author} {\bibfnamefont {L.~F.}\ \bibnamefont {Cugliandolo}},\
  }\href@noop {} {\bibfield  {journal} {\bibinfo  {journal} {Soft Matter}\
  }\textbf {\bibinfo {volume} {7}},\ \bibinfo {pages} {10193} (\bibinfo {year}
  {2011}{\natexlab{b}})}\BibitemShut {NoStop}%
\bibitem [{\citenamefont {Shen}\ and\ \citenamefont {Wolynes}(2004)}]{Shen04}%
  \BibitemOpen
  \bibfield  {author} {\bibinfo {author} {\bibfnamefont {T.}~\bibnamefont
  {Shen}}\ and\ \bibinfo {author} {\bibfnamefont {P.~G.}\ \bibnamefont
  {Wolynes}},\ }\href@noop {} {\bibfield  {journal} {\bibinfo  {journal} {Proc.
  Nac. Acad. Sc. USA}\ }\textbf {\bibinfo {volume} {101}},\ \bibinfo {pages}
  {8547} (\bibinfo {year} {2004})}\BibitemShut {NoStop}%
\bibitem [{\citenamefont {Shen}\ and\ \citenamefont {Wolynes}(2005)}]{Shen05}%
  \BibitemOpen
  \bibfield  {author} {\bibinfo {author} {\bibfnamefont {T.}~\bibnamefont
  {Shen}}\ and\ \bibinfo {author} {\bibfnamefont {P.~G.}\ \bibnamefont
  {Wolynes}},\ }\href@noop {} {\bibfield  {journal} {\bibinfo  {journal} {Phys.
  Rev. E}\ }\textbf {\bibinfo {volume} {72}},\ \bibinfo {pages} {041927}
  (\bibinfo {year} {2005})}\BibitemShut {NoStop}%
\bibitem [{\citenamefont {Wang}\ and\ \citenamefont
  {Wolynes}(2011{\natexlab{a}})}]{Wang11}%
  \BibitemOpen
  \bibfield  {author} {\bibinfo {author} {\bibfnamefont {S.}~\bibnamefont
  {Wang}}\ and\ \bibinfo {author} {\bibfnamefont {P.~G.}\ \bibnamefont
  {Wolynes}},\ }\href@noop {} {\bibfield  {journal} {\bibinfo  {journal} {J.
  Chem. Phys.}\ }\textbf {\bibinfo {volume} {135}},\ \bibinfo {pages} {051101}
  (\bibinfo {year} {2011}{\natexlab{a}})}\BibitemShut {NoStop}%
\bibitem [{\citenamefont {Wang}\ and\ \citenamefont
  {Wolynes}(2011{\natexlab{b}})}]{Wang11b}%
  \BibitemOpen
  \bibfield  {author} {\bibinfo {author} {\bibfnamefont {S.}~\bibnamefont
  {Wang}}\ and\ \bibinfo {author} {\bibfnamefont {P.~G.}\ \bibnamefont
  {Wolynes}},\ }\href@noop {} {\bibfield  {journal} {\bibinfo  {journal} {Proc.
  Nac. Acad. Sc.}\ }\textbf {\bibinfo {volume} {108}},\ \bibinfo {pages}
  {15184} (\bibinfo {year} {2011}{\natexlab{b}})}\BibitemShut {NoStop}%
\bibitem [{\citenamefont {Tailleur}\ and\ \citenamefont
  {Cates}(2009)}]{Tailleur09}%
  \BibitemOpen
  \bibfield  {author} {\bibinfo {author} {\bibfnamefont {J.}~\bibnamefont
  {Tailleur}}\ and\ \bibinfo {author} {\bibfnamefont {M.~E.}\ \bibnamefont
  {Cates}},\ }\href@noop {} {\bibfield  {journal} {\bibinfo  {journal} {EPL}\
  }\textbf {\bibinfo {volume} {86}},\ \bibinfo {pages} {60002} (\bibinfo {year}
  {2009})}\BibitemShut {NoStop}%
\bibitem [{\citenamefont {Szamel}(2014)}]{Szamel14}%
  \BibitemOpen
  \bibfield  {author} {\bibinfo {author} {\bibfnamefont {G.}~\bibnamefont
  {Szamel}},\ }\href@noop {} {\bibfield  {journal} {\bibinfo  {journal} {Phys.
  Rev. E}\ }\textbf {\bibinfo {volume} {90}},\ \bibinfo {pages} {012111}
  (\bibinfo {year} {2014})}\BibitemShut {NoStop}%
\bibitem [{\citenamefont {Van~Kampen}(1981)}]{van1981ito}%
  \BibitemOpen
  \bibfield  {author} {\bibinfo {author} {\bibfnamefont {N.}~\bibnamefont
  {Van~Kampen}},\ }\href@noop {} {\bibfield  {journal} {\bibinfo  {journal}
  {Journal of Statistical Physics}\ }\textbf {\bibinfo {volume} {24}},\
  \bibinfo {pages} {175} (\bibinfo {year} {1981})}\BibitemShut {NoStop}%
\end{thebibliography}%

\end{document}